# Alternative solvents for life: framework for evaluation, current status and future research

William Bains[1,2,*], Janusz J. Petkowski[1,3,4], Sara Seager[1,5,6]

[1] Department of Earth, Atmospheric and Planetary Sciences, Massachusetts Institute of Technology, 77 Massachusetts Avenue, Cambridge, MA, 02139, USA

[2] School of Physics & Astronomy, Cardiff University, 4 The Parade, Cardiff CF24 3AA, UK.

[3] Faculty of Environmental Engineering, Wroclaw University of Science and Technology, 50-370 Wroclaw, Poland

[4] JJ Scientific, Mazowieckie, 02-792 Warsaw, Poland

[5] Department of Physics, Massachusetts Institute of Technology, 77 Massachusetts Avenue., Cambridge, MA 02139, USA

[6] Department of Aeronautics and Astronautics, Massachusetts Institute of Technology, 77 Massachusetts Avenue., Cambridge, MA 02139, USA

* correspondence: William Bains: bains@mit.edu

## *Abstract*

Life is a complex, dynamic chemical system that requires a dense fluid solvent in which to take place. A common assumption is that the most likely solvent for life is liquid water, and some researchers argue that water is the only plausible solvent. However, a persistent theme in astrobiological research postulates that other liquids might be cosmically common, and could be solvents for the chemistry of life. In this paper we present a new framework for the analysis of candidate solvents for life, and deploy this framework to review substances that have been suggested as solvent candidates. We categorize each solvent candidate through four criteria: occurrence, solvation, solute stability and solvent chemical functionality. Our semi-quantitative approach addresses all the requirements for a solvent not only from the point of view of its chemical properties but also from the standpoint of their biochemical function. Only the protonating solvents fulfil all the chemical requirements to be a solvent for life, and of those only water and concentrated sulfuric acid are also likely to be abundant in a rocky planetary context. Among the non-protonating solvents liquid $CO_2$ stands out as a planetary solvent, and its potential as a solvent for life should be explored. We conclude with a discussion of whether it is possible for a biochemistry to change solvents, as an adaptation to radical changes in a planet's environment. Our analysis provides the basis for prioritizing future experimental work exploring potential complex chemistry on other planets.

# 1 Introduction

Life is a complex, dynamic chemical system (National Research Council 2019). The consensus is that attempts to ‘define’ life are futile (e.g. (Cleland 2019, Mariscal and Doolittle 2020)), and there is no consensus on the necessary properties of life. However all descriptions agree that life as we know it is chemical, is dynamic, self-maintaining and self-propagating, and has genetics (and hence is capable of evolution (Luisi 1998), although evolution is a consequence of a coded description in the genes, not a basic property (Bains 2014)). To achieve these characteristics, a living system requires the rapid and efficient chemistry of complex molecules, including polymers, and sufficient chemical diversity in those molecules to allow specificity of chemical interaction. The requirement for rapid chemistry between complex molecules including polymers or macromolecules requires a dense fluid in which to take place (e.g. (Benner et al. 2004, Hoehler et al. 2020)). The medium has to be fluid to allow molecules to react, and it has to be dense so that macromolecules do not aggregate or physically fall out of ‘solution’ (Bains 2004). This is usually stated as the requirement for a liquid, although we will also discuss briefly the idea that supercritical fluids could be solvents for life.

A common assumption is that the most likely solvent for life is liquid water (e.g. (Cockell and Nixon 2012, Meadows and Barnes 2018, Schwieterman et al. 2018, Hallsworth et al. 2021)). The NASA strategy in the search for life ion our solar system also adheres to the same “Follow the Water” philosophy, e.g. in Mars exploration (Hubbard et al. 2002). Some authors argue that water is the only plausible solvent (Pohorille and Pratt 2012), and without liquid water any environment is a priori uninhabitable (e.g. (Hallsworth et al. 2021)). However there has been a persistent theme in astrobiological research that other liquids might be cosmically common, and could be solvents for the chemistry of life, albeit life completely unlike terrestrial life (overviewed in (Ball 2005)). Several surveys of potential solvents have examined the properties of a range of liquids qualitatively (National Research Council 2019) or quantitatively (Schulze-Makuch and Irwin 2004) as potential alternatives to water as solvents for life.

Here we present a new framework for the analysis of candidate solvents for life, and deploy this method to review substances that have been previously suggested as solvent candidates. Unlike previous analyses, our approach starts with the requirements for life, not with the properties of the solvent, and in particular the requirement that large, stable bodies of a solvent be plausible in the context of a realistic planet. We address all the requirements for a solvent, not just single chemical properties, and we do so quantitatively or semi-quantitatively. Our framework takes previous work forward by allowing for the systematic comparison of the plausibility of different solvents.

We begin by summarising the overarching properties that any solvent must have to be plausible as a solvent for life. We then discuss potential solvents that life could use, with respect to the degree to which they might meet the core criteria. We conclude with a discussion of whether it is possible for a biochemistry to change solvents, i.e. undergo solvent replacement, for example as an adaptation to radical changes in a planet’s surface environment.

Out of all of the solvents considered, only the protonating solvents fulfil all the chemical requirements to be a solvent for life, and of those only water and concentrated sulfuric acid are also likely to be abundant on planetary surfaces. Among the non-protonating solvents liquid $CO_2$ stands out as a planetary solvent, and its potential as a solvent for life should be explored.

# 2 *General considerations for a solvent for life.*

If terrestrial life is representative of life elsewhere (an assumption we are forced to make for lack of any other examples), then the solvent for life needs to have four key properties, based on the role that water plays in terrestrial life.

- Occurrence: It must occur and be stable as a liquid on a planetary surface (or in the crust or clouds)
- Solvation: it must selectively solvate compounds, including polymers, but not dissolve everything.
- Solute stability: some of both dissolved and insoluble compounds must be stable in its presence
- Solvent chemical functionality: the solvent should have chemical functionality to enable it to be an active participant in metabolism not just a passive support for it.

We note that the fourth criterion of chemical functionality is based on the observation that water is an integral part of terrestrial biochemistry, and so is not independent of our assumptions about Earth life. We discuss this assumption further below.

Our framework takes the above criteria and expands on them in a comprehensive and semi-quantitative manner.

## **2.1** Occurrence.

The solvent must be stably present in the context of a rocky planet or moon. In principle, a lifeform could generate its own solvent, and most of the water molecules inside a rapidly metabolising bacterial cell come from its own metabolism, not from external solvent (Kreuzer-Martin et al. 2005). However, if the sole source of the solvent was internal metabolism then when the organism's metabolism slowed, due to stress or lack of nutrient or energy, any loss of solvent to reaction, leakage or evaporation would not be replaced, and the organism would dry out. With no external, abiotic source of solvent to re-solvate it the organism would then be permanently dead. An ecosystem of such organisms would therefore be unstable to any large-scale stress. A solvent that was not present on or near the surface of a rocky body in the absence of life is therefore unlikely to be a solvent for life. While this is not an absolute limit, as we discuss in section 5, it is a strong criterion for preferring one solvent over another.

The criterion of occurrence renders implausible three classes of solvents.

The first class of substances excluded are ones which could in principle stably exist under planetary conditions, but which are likely to be extremely rare. For example, elemental mercury (Hg) is liquid under Earth surface conditions (and some conditions on the surface of Venus and Mars). However mercury only comprises ~2 ppb of crustal rocks on average (Canil et al. 2015), mostly present as the sulfide cinnabar (Rytuba 2003), with similar abundances in the Moon (Haskin and Warren 1991) and meteorites (Lauretta et al. 2001). If these abundances are typical of other rocky bodies, then the likelihood of lakes or oceans of liquid mercury is vanishingly small.

The second class of solvents excluded on the criterion of occurrence are solvents that are unlikely to form abiotically under planetary conditions. We discuss a potential exception to this exclusion in Section 5.

The last and most common category are solvents that can form in sufficient abundance, but which are not stable over geological timescales.

Examples of the second and third categories are provided and discussed in detail below and include formamide and hydrogen peroxide, and ammonia, hydrogen sulfide, hydrogen fluoride and methanol respectively.

The occurrence of a solvent must include that the substance is present as a suitable fluid. This means that is must be sufficiently dense, and have a suitably low viscosity (discussed further below and the SI, section S2). The need for a dense fluid is driven by the need for a medium in which molecules of all sizes can move and interact. The solvent must have a low enough viscosity to allow large molecules to diffuse through it and interact. For example, pitch is technically a liquid, but its viscosity is so high at Earth surface temperatures (Edgeworth et al. 1984) that it is an implausible solvent for life. Thus, pitch would only be considered as a plausible solvent for life at a temperature where its viscosity was low enough to allow rapid molecular movement within it (and of course, if it had a plausible abiotic source). (Branscomb and Russell 2019) also suggest that too low a viscosity is incompatible with the non-equilibrium thermodynamics of macromolecular catalysis. They argue that the viscous forces acting on macromolecules in water are much greater than inertial forces and the molecule's motion is 'overdamped' as a result, enabling efficient non-equilibrium chemistry. We note however that many protein molecular motions are much slower than allowed by viscous forces, in some cases on timescales of minutes or hours (e.g. (Genest et al. 2019), reviewed in (Ishima and Torchia 2000, Bai et al. 2012)). Internal rearrangement in the protein is not viscosity limited in these cases. We consider that the low viscosity of a solvent could be 'overcome' (if indeed it needs to be overcome) by adding internal drag functions to macromolecular movement.

Viscosity could also limit biochemistry through limiting diffusion, and hence reaction rates. However viscosity would have to be higher by several orders of magnitude than the highest observed viscosity values that are compatible with life (see SI, section S2), as diffusion in biochemistry is limited by molecular crowding in the cell rather than by bulk fluid viscosity (Weiss 2014).

Solvation, viscosity and stability are all strongly temperature dependent, and so whether a substance is a suitable solvent for life depends on its temperature. Pressure has little effect at the pressures likely to apply to rocky planet surfaces or atmospheres (Bains et al. 2015), with the exception of the case of supercritical fluids discussed below in the context of carbon dioxide.

Here we address the Occurrence criterion in terms of the likelihood of a surface lake or ocean. We have not separately considered subsurface liquids for three reasons.

Firstly, as noted in the next section, it is plausible to suppose that life requires metals, and so life requires access to mineral phases, either from planetary rock or meteoritic material. The chemistry of solvent:mineral interactions is the same whether the liquid is on the surface or not. Similarly, the liquid needs a source, which is either planetary formation, volcanic emission or photochemistry. The first two sources apply equally to the surface and subsurface, the third applies only to the surface.

Secondly, and pragmatically, we will only be able to have evidence for life on exoplanets if that life is on (or very near) the surface. Life in the internal ocean of an Exo-Europa will not be detectable by any conceivable near-future technology.

Lastly, although there are several examples of large bodies of subsurface liquid in the solar system, these are all either rock or water. We discuss liquid rock in the next section as being unsuitable. Liquid water is (almost) unique among common, low melting point substances that have a solid phase that is less dense than its liquid phase, and hence could have an ice layer than floats on a water ocean. A carbon dioxide or sulfur analogue of Europa is therefore very implausible.

We note that the Occurrence criterion is one of likelihood, not absolute exclusion. For several solvents discussed in Section 3 we conclude that their occurrence is improbable, but this does not mean impossible. However the burden of argument for proposing a solvent that 'fails' the occurrence criterion should be to explain how that solvent occurs stably on a planetary surface, rather than simply assuming that it might. Such an explanation would then focus a search for worlds with the characteristics needed to maintain that solvent, to see if they actually exist.

### **2.2** Solvation.

The most obvious property of a solvent is that it has to dissolve molecules, but this dissolution has to be selective. The solvent needs to be able to dissolve a wide range of molecules, including macromolecules and inorganic ions, and to not dissolve other molecules. The need for solvation of some molecules is obvious (Bains 2004, Hoehler et al. 2020); rapid chemical reactions only occur between molecules in solution or in gas phase, and gas phases cannot dissolve macromolecules and do not allow for high concentrations of any, except small, simple molecules. A solid on the other hand does not allow for rapid diffusion, and hence interactions between molecules.

The need for insolubility of some molecules comes from two requirements. First, the structural elements (e.g. cellulose in the cell walls of plants on Earth) must be essentially insoluble in their surrounding solvent if they are to function as cellular barriers and stable supports. Second, insolubility allows for the existence of molecules in which part of the molecule is highly soluble and another part of the molecule is insoluble in the solvent (Tanford 1978), i.e. molecules that are amphipathic. Such characteristics allows for the assembly of nanostructures such as membranes and globular proteins (Dill 1990), driven by solvation properties.

The principle driver of the selective solubility of amphipathic molecules in water is the hydrophobic effect, an example of entropy-driven solvophobic effects seen in solvents with strong hydrogen bonding networks. A solvophobic effect allows for molecules (or components of molecules) to associate without precipitating. This is one example of a class of liquid:liquid phase separations, which are critical to many aspects of terrestrial biochemistry, especially in eukaryotes (McSwiggen et al. 2019, Peeples and Rosen 2021, Musacchio 2022). Solvents that do not show a solvophobic effect can also fail to dissolve some substances and dissolve others, but the differences in solubility result from different physicochemical forces. Thus, for example, amphipaths can have detergent effects in liquid carbon dioxide (van Roosmalen et al. 2004) even though carbon dioxide has no hydrogens to form hydrogen bonds and no permanent dipole.

The solvation of metal ions is another key characteristic of the solvent suitable for life. Terrestrial life absolutely requires metal ions, which play diverse roles in structure, catalysis and redox chemistry (Hughes and Poole 1989, Da Silva and Williams 2001). Polar solvents inherently are more likely to dissolve metal ions, as they can form a polarized solvation shell around the ion, shielding its charge. Solvents that can self-ionize are able to also provide counter-ions to dissolved ions, and in many cases to complex with them increasing their solubility. The ability to not dissolve some inorganic substances is also important for the formation of structural elements, notably silica, calcium carbonate and phosphates for terrestrial life. Selective solvation of inorganic ions is therefore as important as selective solvation of organic molecules.

It is harder to specify which metals are required for life, as there are many examples where different metals can be 'swapped out' for each other in a biological function (e.g. (Hoffman and Petering 1970, Holmquist and Vallee 1974, Sabbioni et al. 1976, Habermann et al. 1983, Price and Morel 1990, Eady 1996, Bock et al. 1999, Decker and Solomon 2005)), and at least a few where a reaction normally catalyzed by a metalloenzyme can be catalyzed by an enzyme that does not require a metal ion at all (e.g. (Berkessel 2001, Corbett and Berger 2010, Blaesi et al. 2018, Genest et al. 2019)).

Solvation is strongly dependent on temperature, so substances that have high boiling points inherently are more likely to fulfill the solvation criterion. Highly polar solvents are therefore favored under the solvation criterion.

## 2.3 Solute stability.

Some (but not all) molecules, including some that the solvent dissolves, must be stable in the presence of the solvent (Hoehler et al. 2020). By 'stable' we mean that the core structure of the solute remains unaffected by the solvent on a timescale of minutes to years, depending on the role of the metabolite. (This definition does not include transient changes that do not alter the backbone structure of a molecule, such as the reversible protonation of carbonyls in sulfuric acid (Seager et al. 2023)). The stability of solutes dissolved in a solvent depends on temperature, so we consider whether there is a temperature at which a solvent is liquid and at which diverse chemicals are stable. Almost no terrestrial biochemicals are stable in water above 300°C (Daniel et al. 2004, Brunner 2009, Bains et al. 2015, Yakaboylu et al. 2015), but this value does not limit water's potential as a solvent because planetary conditions at which water is a liquid exist substantially below this temperature.

We should note here that stability, as well as solubility criteria discussed above, applies to all molecules, not just ones known to be components of terrestrial biochemistry. For example, sulfuric acid has previously been discounted as a solvent for life because of the well-known instability of sugars to aggressive dehydration in concentrated sulfuric acid (e.g. (Pines et al. 2012)). However, this does not mean that all organic chemicals are unstable in concentrated sulfuric acid, as we discuss below (Seager et al. 2023, Spacek et al. 2023, Petkowski et al. 2024, Seager et al. 2024a, Seager et al. 2024b)

The requirement for chemical stability of molecules dissolved in a solvent provides an absolute upper limit on the temperature for chemical life. Liquid rock is extremely common in the core of terrestrial planets, and even on the surface of some exoplanets (Winn et al. 2018), and has been speculatively suggested as a solvent for life (Feinberg and Shapiro 1980). However, at temperature of thousands of Kelvin almost all molecules rapidly degrade from thermal breakdown, regardless of their chemistry. The gases outgassed by volcanoes are representative of molecules that are stable at liquid rock temperatures, and show an exponential decrease in abundance with molecular size. Even quite small molecules like propane and thiophene are vanishingly rare in volcanic gases at 200°C (Tassi et al. 2009, Tassi et al. 2010), and are likely to be even rarer in the rock itself. Consequently, liquid rock does not support sufficient solute stability and is an unlikely solvent for a biochemistry.

## 2.4 Solute stability and diversity.

The chemistry of life requires diversity, in two respects: stability and function.

First, life requires molecules that differ in chemical stability at ambient temperature. Some metabolic intermediates have to be marginally stable on a timescale of days, but be able to react readily. Thus adenosine triphosphate (ATP) hydrolyses slowly in water at temperatures at which life can live (Kunio 2000, Daniel et al. 2004), but DNA is stable for millennia. Excessive instability is undesirable as molecules would degrade before biochemistry can use them. Excessive stability is also undesirable, as illustrated by elemental nitrogen ($N_2$). While in principle the reduction of $N_2$ by nicotinamide adenine dinucleotide (NADH) to form $NH_3$ is highly exergonic, the extreme strength of the nitrogen-nitrogen triple bond requires life to use an additional 8 moles of ATP per mole of $NH_3$ formed to overcome the activation energy of this reaction (Kim and Rees 1994). The solvent may play a role in the relative stability of metabolites; if the solvent is chemically inert, the requirement for a range of stabilities of molecules places additional constraints on the chemistry from which biochemistry can be constructed.

Second, both stability and solvation have to allow for the existence of a diverse chemical space to provide the chemical functionality needed for life. In "Chemical space", relevant properties of a large set of compounds are used to map them onto a space that can be used to predict the properties or function of other molecules (e.g. (Dobson 2004, Kirkpatrick and Ellis 2004, van Deursen and Reymond 2007, Reymond et al. 2010)). The chemical space accessed by life must be structurally and functionally diverse (Hoehler et al. 2020, Bains et al. 2021b). For example, perfluorocarbons are stable to temperatures approaching that of the surface of Venus (Lewis and Naylor 1947, Logothetis 1989), but a metabolism made entirely of perfluorocarbons is implausible because of the notorious chemical homogeneity and inertness of these substances that do not provide sufficient chemical functionality needed for life. One specific example of the chemical functionality of life is the chemical functionality of the solvent itself, which we discuss in the next section.

## 2.5 Solvent chemical functionality.

The solvent for life can in principle participate in biochemistry as a reactant as well as a solvent. Terrestrial life uses water as a reagent in energy capture and transfer, in polymerization and depolymerization of large molecules, in forming proton gradients that power diverse cellular processes, and in many other contexts. Water's ability to donate and accept protons enables stable charges on amino and carboxylate groups, themselves central to interacting with inorganic ions. Protonation is also central to a wide range of catalytic mechanisms, and proton gradients are used in a range of energy handling processes such as oxidative energy metabolism.

We note that the need for chemical functionality in the solvent is not driven by fundamental physical principles, but rather is an extrapolation from the role of water in Earth life. The biochemistry of Earth provides a precedent for suggesting that it may be possible to construct a biochemistry that does not rely on solvent chemical functionality. Many classes of metabolic reactions involve functional groups that are not stable in water, and so cannot be transferred using water as a reagent in their chemistry. Examples include electron transfer (redox reactions) and the wide range of transferase reactions transferring chemical groups such as methyl between molecules. In these cases, life uses carrier molecules for the functional groups. In the case of electron transfer, carriers include $NAD^+$/NADH, iron-sulfur proteins, ubiquinones and others, in the case of one-carbon fragments carriers include folate and S-adenosylmethionine (SAM). If the solvent for life did not provide chemical function beyond solvation, then metabolism would have to include carrier molecules for a wider range of reactions than terrestrial biochemistry. This would make biochemistry more complex, but not impossible. It therefore seems reasonable that a solvent that can participate in biochemistry in this way is more plausible as a candidate solvent for life than a solvent that is just a passive support for reacting molecules. We note however that the lack of solvent chemical functionality may not be an absolute barrier to a solvent's use in biochemistry.

Furthermore, chemical functionality of water can also be detrimental for life as it also reduces the stability of biomolecules. Terrestrial life requires multiple layers of repair enzymes to repair the damage done to nucleic acids by spontaneous hydrolysis (Shapiro 1981, Gates 2009, Huang and Zhou 2021). Thus, even if chemical functionality is required by life on Earth, the amount of chemical functionality is not clearly delimited by terrestrial biochemistry.

We conclude that the criterion for solvent chemical functionality is a less stringent constraint on the solvent for life than the other three constraints.

## 2.6 Origin of life

We have not discussed the criteria of a solvent necessary for the origin of life (OOL), only for life's continued operation. We exclude consideration of OOL because there is no consensus on many fundamental aspects of the origin of life even on Earth, including no consensus on the environment in which it started (Bains 2020). This extends to disagreement on such basic aspects as the solvent in which life's chemistry originated (Schoffstall and Laing 1985, Benner et al. 2004, Sydow et al. 2017, Ziegler and Benner 2018, Gull et al. 2023, Sydow et al. 2023). Exploration of origins scenarios for the solvents discussed here remains work for the future.

We now discuss the solvents that have been considered for life in more detail, with reference to how they fulfil the four criteria overviewed above.

# 3 *Candidate solvents for life*

## 3.1 Liquid Water

It is indubitable that water is a good solvent for biochemistry. Its properties have been extensively reviewed elsewhere (Pohorille and Pratt 2012, Hoehler et al. 2020), and will only be briefly summarised here.

### 3.1.1 Liquid Water: Occurrence

Water is cosmically abundant. It is composed of the first and third most abundant element in the Universe, and is by far the most abundant gas emitted by terrestrial volcanoes (see SI, Section S1). Its photochemical breakdown to H and OH is only a net loss to the planet if the H is subsequently lost to space (as has been proposed was the fate of water on Mars (Jakosky 2021), and potentially Venus (Way et al. 2016)). On Earth, water lost to space could be effectively replaced through volcanism. This abundance and stability means that water is by far the most common liquid (other than magmas) in rocky solar system bodies, being present in liquid form on the surface of the Earth, on the surface of early Mars and possibly Venus, and in the interior of a number of the larger icy moons.

Obviously, water's viscosity enables biochemistry to happen at all the temperatures at which water is a liquid on Earth. This sets conservative limits on the viscosity of candidate solvents for life, discussed in more detail in the SI, Section S2.

Water also shows little change in its physical, chemical and solvation properties with temperature (Pohorille and Pratt 2012). This allows a single biochemistry to function in water over a wide range of temperatures, which means that water is a suitable solvent for life over that range of temperatures (compare this characteristic to the substantial changes in physical properties of liquid sulfur or supercritical $CO_2$ over relatively small changes in temperature, discussed below). On the criterion of Occurrence water is one of the strongest candidate solvents for life in our list.

### 3.1.2 Liquid Water: Solvation

Water is highly favored as a solvent for life because as well as being an excellent solvent for polar molecules and salts, many molecules are highly *insoluble* in water, allowing for a 'hydrophobic' force that drives the folding of proteins, assembly and stability of membranes etc. (Tanford 1978, Pratt and Pohorille 1992, Pohorille and Pratt 2012). However this is not a unique property of water. Of the solvents considered, only water, ammonia, hydrogen fluoride and sulfuric acid have the dense hydrogen bonding networks in the liquid state needed drive a solvophobic effect, and micelle formation has been observed in all four liquids (discussed below).

The ability of protonating solvents – paradigmatically water – to enable proteins and nucleic acids to adopt the unique, functional 3D structures essential for their activity is considered as prima facie evidence that non-protonating solvents cannot be solvents for life. This is even true of intrinsically disordered proteins, which have to be poised to adopt ordered structure as required. The ability of water to satisfy the solvation needs of life need not be discussed further.

#### 3.1.3 Liquid Water: Solute Stability

An enormous range of chemicals are stable in solution in cold water, but stability depends on temperature, and supercritical water is widely used as a method to comprehensively break down complex organic materials (Daniel et al. 2004, Brunner 2009, Bains et al. 2015, Yakaboylu et al. 2015). As with all solvents, stability of dissolved molecules depends on temperature. However, water can exist as a liquid at temperatures where many organic chemicals are stable.

Water's example illustrates the range of stability that life requires, or can tolerate. This includes the very short, almost transient water stability of certain crucial metabolites to the very long lifetimes of some molecules that are essentially completely water-stable. For example, carbamoyl phosphate, a key metabolite in the urea cycle, has a half-life to hydrolysis of between 0.5 and 1.5 seconds at 122°C (Bains et al. 2015), the temperature at which *Methanopyrus kandleri* can grow (Takai et al. 2008), and NADH, a central enzyme cofactor in metabolism, has a half-life of less than 10 minutes. On the other hand DNA in bone is hydrolysed at a rate of around $5.5 \cdot 10^{-6}$ bases/year (Allentoft et al. 2012), and DNA in bacterial spores can remain sufficiently intact to support life for 20 million years or more (Cano and Borucki 1995). Stability therefore requires that a substantial chemical space be stable over a range of timescales. Liquid water provides this range of stabilities.

#### 3.1.4 Liquid Water: Solvent Chemical Functionality

Water is also an essential reagent in terrestrial biochemistry, and as such water is not just a solvent but is a participant in biochemistry. It is widely stated that water's chemical properties are important in the chemistry of life, and specifically for selectivity at the origin of life (National Research Council 2019, Hoehler et al. 2020). Moreover, water's ability to solvate protons and hydroxyl ions is central to a wide range of biochemistry, including proton gradient-based bioenergetics. We have touched on the chemical role of water in biochemistry above, and will not elaborate on it further here.

#### 3.1.5 Liquid Water: Conclusion

It need not be stated that water is a solvent for biochemistry. However, the analysis above suggests that it is indeed an optimal solvent for all four of the criteria presented in this paper.

### 3.2 Liquid Ammonia

Liquid ammonia is one of the most widely mentioned alternative solvents to water (e.g. (Feinberg and Shapiro 1980, National Research Council 2019)). Ammonia as a solvent should be distinguished from ammonia solution in water. The chemistry of pure ammonia is quite different from that of an aqueous solution of ammonia. Here we discuss pure or nearly pure liquid ammonia as a solvent. We conclude that ammonia fails the Occurrence criterion for a solvent for life.

### 3.2.1 Liquid Ammonia: Occurrence

Ammonia is unlikely to fulfil the Occurrence criterion, for two reasons.

Firstly, ammonia is photochemically labile, being broken down to form $H_2$ and $N_2$ as photolysis end-products (Strobel 1975, Huang et al. 2022a). $N_2$ is extremely stable and unreactive, unlike OH and $O_2$, and so the nitrogen from photochemically broken ammonia is permanently trapped as $N_2$. In the absence of life, ammonia is only observed in a planetary context in giant planets, where $NH_3$ is regenerated from $N_2$ and $H_2$ at high temperatures and pressures deep in the atmosphere (Moeckel et al. 2023). Ammonia can be maintained in the atmosphere of a rocky planet if it is regenerated by life (Seager et al. 2013, Huang et al. 2022a), but as discussed above, life that relies solely on ammonia as a solvent that is regenerated in this way is walking a perilous ecological tightrope.

The photochemical lability of ammonia is not a direct issue for a subsurface ocean such as that under the surface of the Jupiter's moon, Europa. However, a Europa-style moon can only form with a water/ammonia ocean if the ocean is at least 80% water. The density of solid ammonia and of the water ammonia eutectic are greater than those of the corresponding liquid phases up to at least 2000 bar (Hogenboom et al. 1994). An "ammonia Europa" would therefore tend to lose liquid ammonia from its subsurface ocean to the surface over geological time, there to be irreversibly removed by photolysis. An ammonia/water ocean below an ammonia/ice shell is only stable with >~40% water in the mixture, the chemistry of which is more similar to that of water than that of anhydrous liquid ammonia.

The photochemical lability of ammonia is mirrored by its thermal lability. Ammonia is rarely present in volcanic gases, and when detected there (probably at least in part from biogenic sources) it is present at an abundance that is at least 10-fold lower than $N_2$ gas. Ammonia is readily broken down by thermal processing, even in a high pressure, strongly reducing environment (Larson and Dodge 1923). Therefore even in the absence of photochemical breakdown, volcanic or tectonic processing will reduce the ammonia content of an ammonia water ocean.

Secondly, ammonia and water are expected to co-occur on planets and moons, and water is expected to be more abundant due to both the higher cosmic elemental abundance of oxygen over nitrogen (and hence higher abundance of water over ammonia in interstellar clouds (Wooden et al. 2004)) and due to the photochemical lability of ammonia as compared to water in environments exposed to UV radiation. As water and ammonia are completely mutually miscible (Leliwa-Kopystyński et al. 2002), any ammonia will end up in an ammonia-water mixture, probably one dominated by water, and the chemistry of such a mixture will essentially be the chemistry of alkaline water. Freezing an ammonia-water mixture cannot generate pure ammonia, as ammonia hydrates melt at a lower temperature than ammonia itself (Chua et al. 2023). In principle a mixture of >80% ammonia, <20% water could be frozen to generate pure ammonia ice, the remaining eutectic liquid could then be removed before it froze, and the ammonia ice thawed to create nearly anhydrous liquid ammonia (a scenario similar to one postulated for formamide below). However, this is a highly contrived scenario and unlikely to happen outside the laboratory. Thus, although a sufficiently cold, massive rocky planet might retain its ammonia against photolysis over geological time, that ammonia is still very unlikely to be a solvent in its own right, rather than a solute in water oceans.

In conclusion, while ammonia may play a bigger role in an exoplanet biochemistry than it does on Earth, as speculated by (Seager et al. 2013, Huang et al. 2022a), it is unlikely to be a solvent in its own right (Seager et al. 2013, Bains et al. 2014, Huang et al. 2022a). Protecting the ammonia from photolysis in a sub-surface ocean or on the surface of a 'rogue planet' (Scholz et al. 2022) would make ammonia as a solvent more plausible, but to be a solvent in its own right ammonia would have to be at least 10 times as abundant as water, which seems implausible. We note that in such an environment liquid ammonia would have similar viscosity to water at 100$^{o}$C, which is known to be compatible with life (see SI, Section 2).

### 3.2.2 Liquid Ammonia: Solvation

As a solvent ammonia is almost as good a solvent for life as water. It has a high dipole, a hydrogen bonding network that facilitates solvophobic interactions and the structures they enable (Griffin et al. 2015). It is a good solvent for a wide range of compounds including inorganic salts (Hunt 1932), and has similar chemical reactivity as water (Franklin 1905, Griffin et al. 2015, National Research Council 2019). A wide range of nitrogen analogues of terrestrial biochemicals than might be favored in an ammonia solvent or an ammonia-rich environment have been suggested (National Research Council 2019).

### 3.2.3 Liquid Ammonia: Solute Stability

In general, solutes are more stable in liquid ammonia than in water simply because liquid ammonia is colder than water, with a freezing point of -77$^{o}$C and a boiling point of -33$^{o}$C at 1 bar. Ammonolysis reactions are well known, but refer to breakdown of substances in aqueous ammonia solutions rather than in pure ammonia (Stevenson 1948).

The chemical diversity available to chemistry in liquid ammonia is higher than that in water. Classes of compounds that are unstable to hydrolysis by water, such as silanes, germanes and arsanes, are stable in liquid ammonia (Fernelius and Bowman 1940), so the chemical space available to life in liquid ammonia would be larger than that available to life in water. Compounds that would be extremely unstable at terrestrial temperatures, such as diazomethane and azidomethane, are potentially stable at the temperature of liquid ammonia (or ammonia/water mixtures) (Raulin et al. 1995). Nitrogen-containing analogues of phosphates and carbonyl groups are known and likely to be favored over their oxygen equivalents in liquid ammonia (Aspinall et al. 2002, Bains 2004, Benner et al. 2004). Ammonia analogues of peptides and sugars are known or have been plausibly modelled (Firsoff 1963, Raulin et al. 1995, Oliveira et al. 2014), and nucleic acids with one or more oxygen replaced by an NH group are well known from terrestrial biology and pharmacology (Oliveira et al. 2014). The idea of an ammonia analogue of terrestrial biochemistry is therefore well established.

The stable solvation of metals in liquid ammonia allows formation of organometallic compounds that would be extremely reactive in water, such as alkyl sodium and potassium compounds (Kraus 1940). The relatively facile reaction of these chemicals with each other and with other compounds to make, rearrange and break carbon-carbon and carbon-heteroatom bonds could compensate for the inherently slower reactivity of all chemistry at liquid ammonia temperatures.

### 3.2.4 Liquid Ammonia: Solvent Chemical Functionality

Ammonia is a protonating, hydrogen bonding solvent like water, but also has significant chemical differences from water. Ammonia does not solvate protons as well as water, and dry ammonia does not ionize to the same degree ($K_{eq}$ of the reaction $NH_3 \leftrightarrow NH_4^+ + NH_2^-$ is ~$10^{-33}$ at 220 K (Greenwood and Earnshaw 1997)). In dry ammonia proton gradients are therefore less likely than in water, but electron gradients could be formed instead, as dry liquid ammonia solvates electrons and elemental alkali metals (Greenwood and Earnshaw 1997). The ability to solvate electrons makes liquid ammonia a much more flexible solvent for redox chemistry than water (Lagowski 2007). The solvation of electrons as $NH_2^-$ ions raises the intriguing idea that bioenergetics in liquid ammonia could be powered by electron gradients rather than proton gradients.

### 3.2.5 Liquid Ammonia: Conclusion

We conclude that ammonia is an excellent candidate solvent for life on the criteria of Solvation, Stability and Chemical Functionality. However, it fails as a plausible candidate solvent on Occurrence, as liquid ammonia is unlikely to be present on a planetary surface as a solvent in its own right distinct from an aqueous solution of ammonia in water.

## 3.3 Concentrated Sulfuric Acid

Interest in sulfuric acid as a solvent for life has been stimulated by the possibility of life in the clouds of Venus (e.g. (Dartnell et al. 2015, Limaye et al. 2018, Bains et al. 2021a, Limaye et al. 2021, Mogul et al. 2021, Seager et al. 2021, Patel et al. 2022, Bains et al. 2023)).

We note that the solvent under consideration here is concentrated sulfuric acid, i.e. largely pure sulfuric acid, not sulfuric acid solutions in water. The chemistry of 98% w/w sulfuric acid is fundamentally different to 80% w/w acid; the former acts as a distinct solvent with unique chemical properties, the latter as very concentrated solution of an acid in water (see SI, Section S3). Here we discuss concentrated sulfuric acid only, as a chemically distinct solvent species. We note that an environment might contain sulfuric acids with differing amounts of water, as is true of the clouds of Venus which vary from 81% - 100% acid (w/w) (Hallsworth et al. 2021, Bains et al. 2023). If life were to use sulfuric acid as a solvent in such an environment, then it would have to adapt to the differing chemical properties of 80% w/w and 100% w/w sulfuric acid. The primary difference between the two is in the spectrum of molecules that are stable to solvolysis. We address this potential challenge further below.

### 3.3.1 Sulfuric Acid: Occurrence

(Ballesteros et al. 2019) predicted that sulfuric acid could be a common liquid on exoplanets, where it would be formed from photochemical oxidation and subsequent hydration of volcanic $SO_2$ (Titov et al. 2018), or directly emitted by volcanoes (Zelenski et al. 2015). In very water-poor planetary environments, the result would be an accumulation of concentrated sulfuric acid rather than dilute acid in aqueous solution.

The example of sulfuric acid illustrates that the stability of a solvent and its eventual abundance depends on its planetary environment, for two reasons.

Firstly, sulfuric acid is extremely hygroscopic, and water and sulfuric acid are completely mutually miscible. Therefore, concentrated sulfuric acid will only accumulate as a solvent in its own right (rather than as a solute in water) in an environment that is very substantially depleted of water. If a planet loses almost all its water, sulfuric acid could come to be the dominant liquid. Such a scenario has been suggested for Venus, and may occur on worlds orbiting in the 'habitable zone' of mature red dwarf stars if their orbits complete any planetary migration before the star cools to its equilibrium, fusion-supported temperature. Such worlds could be common around low mass red dwarf (M dwarf) stars. Such planets, that are initially too hot to allow for surface liquid, are sufficiently cool after tens or hundreds of millions of years and could fall into the Belatedly Habitable Zone of that star (Tuchow and Wright 2023). The Kelvin-Helmholtz thermal radiation from a pre-main sequence red dwarf can be 1000 times that of its steady state main sequence luminosity (Hayashi and Nakano 1963), which means any planet close enough to the star to be in the habitable zone when the star is at its main sequence luminosity will have been baked dry in the first few tens to hundreds of millions of years of the star's life. Subsequent volcanic emission of $SO_2$ and $H_2SO_4$ would then generate sulfuric acid which could condense into surface liquid. Sulfuric acid's abundance in planets around M-dwarfs will depend on the relative timing of planetary migration compared to cooling of the primary. The occurrence of sulfuric acid on planets around sun-like stars will depend on how common planets like Venus are (Arney and Kane 2020).

Another scenario for dehydrating a small planetary body is through tidal heating, as illustrated by the Jupiter's moon Io. Io's surface 'ices' are almost exclusively sulfur compounds, primarily $SO_2$ and elemental sulfur, with only trace amounts of $H_2S$ and $H_2O$ (Nash and Betts 1998). With surface temperatures ranging from -143°C to -130°C (Spohn et al. 2014) Io is too cold for even 94% w/w sulfuric acid to be a liquid. However a larger satellite (sufficient to retain a 6 millibar atmosphere like Mars) orbiting a planet that was nearer to its host star than Jupiter (and hence had a higher black body surface temperature of >-33°C could evolve to become a tidally heated super-Io, which could then support surface (or subsurface) liquid sulfuric acid.

The scenario envisaged here is different from any that could give rise to a water-depleted ammonia ocean. The water-stripping mechanisms that apparently operated on Venus have not depleted atmospheric and crustal sulfur reservoirs from which sulfuric acid can be generated. By contrast, any process that depletes an atmosphere and crust of water is likely to also deplete it of ammonia. Thus, a 'dry ammonia' world is unlikely, while a 'dry sulfuric acid' world is not only plausible but has Solar System precedent.

Secondly, sulfuric acid will probably react with some minerals. Weathering of rocks by aqueous sulfuric acid solutions is well known (e.g. (Golden et al. 2005, Hausrath et al. 2013)). How concentrated sulfuric acid interacts with minerals has not been explored, but it seems likely that reaction will occur with some minerals. The unknown is how much: after all, water weathers rocks, but this does not preclude the presence of oceans on Earth. Therefore, it is unknown whether stable surface lakes or oceans of sulfuric acid could exist on a 'dry' planet, but possibly they can if the surface layers of the crust are fully converted to sulfates, or are made of minerals resistant to attack by concentrated sulfuric acid, such as silica

Sulfuric acid can exist stably as clouds which do not come into contact with the surface. Again, Venus provides an example of an environment in which clouds of sulfuric acid have probably existed for geological time.

Sulfuric acid is more viscous than water at any given temperature. Comparison of the viscosity of sulfuric acid as a function of temperature shows that above ~30°C its viscosity is less than the maximum seen in the cytoplasm of some Earth life (see SI, Section S2), and so above 30°C sulfuric acid is a plausible solvent on viscosity grounds. If the 30°C is a genuine lower limit for sulfuric acid as a solvent for viscosity reasons, then if temperature dropped below 30°C a sulfuric acid-based organism's cytoplasm would effectively 'vitrify', as does water in supercooled terrestrial organisms. This change would not necessarily kill the organisms, but they would have to be warmed up again to continue replication. However, it is not firmly established that more viscous solutions cannot allow biochemistry.

We conclude that liquid sulfuric acid fulfils the occurrence criterion for clouds, and probably fulfils it for surface liquid, but that further work on rates of reactions with plausible surface rocks is needed to confirm this conclusion.

#### 3.3.2 Sulfuric Acid: Solvation

Recent modelling (Bains et al. 2021b) and experimental studies (Seager et al. 2023, Spacek et al. 2023, Petkowski et al. 2024, Seager et al. 2024b) have shown that sulfuric acid can stably dissolve a wide range of substances, including the bases from DNA and many biological amino acids. Extensive research shows that diverse surfactants can form micelles, and may be able to form liposome-like structures, in concentrated sulfuric acid (McCulloch 1946, Miron and Lee 1963, Steigman and Shane 1965, Menger and Jerkunica 1979, Müller and Miethchen 1988, Müller 1991a, Müller 1991b, Müller 1991c, Müller and Giersberg 1991, Müller and Giersberg 1992, Müller and Burchard 1995, Torn and Nathanson 2002). Some of these are compounds that are uniquely stable in concentrated sulfuric acid, such as carbonium compounds (Menger and Jerkunica 1979). Interestingly, water is found to be a chaotrope in concentrated sulfuric acid, disrupting fatty acid micelles (Steigman and Shane 1965), illustrating the unexpected and complex behaviour of this liquid. A range of polymers are known that are stable in sulfuric acid, some of which are soluble, others insoluble (Bains et al. 2021b). Sulfuric acid also dissolves a wide variety of metal ions. We conclude, as summarized before (Bains et al. 2021b), that sulfuric acid has good solvation properties for life.

#### 3.3.3 Sulfuric Acid: Solute Stability

The physical and chemical properties of sulfuric acid change dramatically with the addition of 20% w/w water (Liler 2012), and hence the stability of solutes in that acid change with acid concentration as well. 80% w/w sulfuric acid acts like a very strong aqueous acid, its chemistry dominated by $H_2O$ and $H_3O^+$. Sulfuric acid of >90% w/w concentration acts as an oxidizing, protonating solvent, with very different reaction mechanisms. (We note that this is different from the ammonia example above, where even 1% of water in ammonia dramatically changes its ability to perform some chemistries such as dissolving alkali metals.) The consequence of these characteristics is that reaction kinetics can change in a complex way as acid concentration increases, and some substances can actually be more stable in concentrated sulfuric acid than in dilute acid (See examples in the SI, Section S3). Many components of terrestrial biochemistry would be rapidly and completely destroyed in pure sulfuric acid solvent, notably any biochemicals containing sugar moieties such as ATP, NADH, or RNA, and so any sulfuric-acid-based biochemistry would be completely different from terrestrial water-based biochemistry.

However, a range of diverse chemicals are predicted to be stable in sulfuric acid at concentrations of >90% (Bains et al. 2021b), and reactive organics have been found to form specific stable products in sulfuric acid (Spacek et al. 2023), and not 'tar' as might be expected. Surprisingly, some components of terrestrial biochemistry are stable for periods of months in 81% w/w and 98% w/w sulfuric acid, including DNA bases (Seager et al. 2023) and 19 of the 20 proteinaceous amino acids (Seager et al. 2024a). These include glutamine, serine and cysteine as well as hydrophobic amino acids like valine, alanine and leucine, showing that amino acids with chemically diverse side-chains are stable, although some side-chains are reversibly modified by sulfation. Diverse polymers are stable in sulfuric acid (reviewed in (Bains et al. 2021b)), some of which are soluble, some not. Carbonyl groups are readily protonated in concentrated sulfuric acid, and as a result some carbonyl compounds such as aldehydes are highly reactive. The carbonyl group is central to terrestrial biochemistry. However Benner has pointed out the vinyl group ($C=CH_2$) has similar reactivity in concentrated sulfuric acid as the carbonyl group ($C=O$) has in water (Benner et al. 2004), and so could provide equivalent function.

If an environment contains a range of sulfuric acid concentrations, then organisms could adapt to this in either of two ways. It could use biochemistry that was appropriately stable in all concentrations of sulfuric acid, or it could adopt a dormant, protective form in lower concentration acid and only maintain active metabolism in concentrated acid >95%. The later of these hypothetical strategies is analogous to terrestrial organisms that adapt to transiently low water environments by forming dormant forms, which can then germinate and return to active metabolism when the water activity rises (Clegg 2001). The former would restrict the chemical space available to life, as only a subset of molecules that are stable in 98% w/w sulfuric acid are also stable in sulfuric acid containing significant water (See SI Section 3 for some examples of this effect, and further discussion in Section 5 below).

We conclude that diverse chemistry is stable in sulfuric acid. However, sulfuric acid is an aggressive solvent reactivity of which towards organics has not been systematically explored, so whether sufficient chemical diversity exists to support a biochemistry has not yet been experimentally demonstrated.

#### 3.3.4 Sulfuric Acid: Solvent Chemical Functionality

Sulfuric acid is a polar, protonating solvent like water, with a strong hydrogen bonding network and a dipole moment similar to water. Sulfuric acid provides as much chemical functionality as water. The solvent self-ionizes to form positive and negative ions $HSO_4^-$ and $H_3SO_4^+$ (Cox 1974) more readily than water forms $OH^-$ and $H_3O^+$ (~$10^{-3}$ M ions in 100% acid, vs $10^{-7}$ M ions in 100% water at room temperature), which can provide ions for charge neutralization, ion solvation, polymer stabilization and charge separation for energy capture. Concentrated sulfuric acid shows solvophobic effects, as noted above. It can reversibly sulfate a range of chemical functional groups such as alcohols and thiols, and can drive reversible protonation of aromatic systems (as well as sulfonation and degradation reactions). Sulfuric acid therefore has a similar potential to participate in biochemistry to water.

### 3.3.5 Sulfuric Acid: Conclusion

We conclude that sulfuric acid is a surprisingly plausible candidate solvent for life on the criteria of Solvation, Chemical Functionality and Occurrence. On the criterion of Stability it presents some challenges, as it is quite reactive, but preliminary work suggests that it can satisfy this criterion as well.

## 3.4 Liquid Sulfur

Liquid sulfur is one of only four liquids naturally occurring on or above the surfaces of rocky or icy solar system bodies (the others being water, sulfuric acid and methane/ethane). Its astrobiological potential has not been explored, and little data is available, so we will consider it only briefly, with an emphasis for the need for future research on this fascinating planetary liquid.

### 3.4.1 Liquid Sulfur: Occurrence

Elemental sulfur is the third most common liquid erupted by volcanoes (after water and magma) on Earth (Mora Amador et al. 2019), and can form extended pools of liquid sulfur on the surface (Oppenheimer and Stevenson 1989) or below a water layer (Takano et al. 1994, De Ronde et al. 2015, Sydow et al. 2017, Malyshev and Malysheva 2023) in subaerial and submarine volcanoes. Liquid sulfur is also erupting on Io (Schneider and Spencer 2023), and could be present in volcanic systems on Venus. However, such volcanic systems are problematic as the basis for sulfur-based life, for two reasons.

Firstly, they are transient, so at least life based in volcanic liquid sulfur would have to consist exclusively of organisms that were occasionally 'wetted' with solvent, rather than being bathed in it continuously. While this is not a complete barrier to life, as illustrated by the existence of desert flora on Earth that are only occasionally wetted by water, it makes the habitat more perilous. Transient volcanic sources of liquid sulfur might include subsurface reservoirs, but terrestrial precedent suggests that subsurface reservoirs of liquid sulfur do not persist over geological time.

Secondly, liquid sulfur's properties vary substantially with temperature. Between 120°C and ~155°C it consists mainly of $S_8$ rings, but above 155°C the predominant molecular species changes to polymeric chains. As a result the viscosity of the liquid increases more than 10,000-fold, dielectric constant increases, and other chemical properties change significantly (Powell and Eyring 1943). Such dramatic changes in liquid sulfur's properties mean that in reality it is effectively composed of "two solvents" that interchange over a relatively narrow temperature range. Specifically, the viscosity of sulfur above ~170°C far exceeds that in which any biochemistry is observed to occur on Earth. We note that quite small amounts of impurity can change this behaviour dramatically. For example, the addition of 0.25% elemental iodine can reduce the viscosity of liquid sulfur at 180°C 1000-fold (Powell and Eyring 1943). The presence of such impurities however is still insufficient to bring the viscosity of liquid sulfur within the ranges seen in terrestrial organisms (See SI, Section S2 for more on viscosity constraints). This complex temperature-dependent behaviour makes liquid sulfur an effective solvent for life only between 115°C and ~170°C, a narrow temperature range for volcanic systems. The only plausible habitat in which liquid sulfur would be present for extended periods would be a planetary surface that was between this narrow temperature range. Such surface conditions would have to be also free of oxidants or reductants that would convert sulfur to $SO_2$ or $H_2S$ respectively. Liquid sulfur also reacts with water at >120°C (Ellis and Giggenbach 1971), in the reaction whose overall stoichiometry is

$$4S + 4H_2O \rightarrow 3H_2S + H_2SO_4$$

so an environment hosting liquid sulfur as a solvent would also have to be extremely dry.

In summary we conclude that the Occurrence criterion is hard to meet for liquid sulfur, but does not rule it out.

#### 3.4.2 Liquid Sulfur: Solvation

Almost nothing is published on the solubility of organic materials in liquid sulfur. The only published material is on the solubility of hydrogen sulfide and sulfur dioxide (Fanelli 1949, Touro and Wiewiorowski 1966, Marriott et al. 2008). Sulfur is not a polar or a protonating solvent, and so would be expected to be a hydrophobic solvent and to dissolve polar molecules poorly. However calcogenic metals complex sulfur very well, (e.g. (Dravnieks 1951)), and elements such as As, Sb, Se, Te, Hg and Cu are found to be concentrated in fumarolic sulfur (Shevko et al. 2018). Beyond this we cannot speculate.

#### 3.4.3 Liquid Sulfur: Solute Stability

Solute stability in liquid sulfur has not been explored. Anecdotally solidified sulfur from volcanic liquids is almost always coated in black precipitate, suggesting reaction with environmental material (e.g. (Sydow et al. 2017, Malyshev and Malysheva 2023)). Liquid sulfur contains open chains at all temperatures, with sulfur radicals or ions at the ends, which would be expected to be reactive towards all molecules dissolved in it. All chemical reactions happen faster with elevated temperature, so the high melting point of sulfur mitigates against the potential thermal stability of organic compounds dissolved in it. This combination of factors leads us to expect that few classes of organic compounds will be stable in liquid sulfur. However, we thought the same of concentrated sulfuric acid, and have been proven wrong, so the expectation of the instability of organic compounds in liquid sulfur awaits experimental verification.

#### 3.4.4 Liquid Sulfur: Solvent Chemical Functionality

Liquid sulfur is quite chemically active, as noted above. It is likely to be a good mediator of redox reactions, as sulfur can stably exist in many redox states. Sulfur can also mediate reductive photochemistry (Li et al. 2022). However, the chemistry of liquid sulfur has not been explored widely, so we cannot draw solid conclusions on its chemical functionality.

#### 3.4.5 Liquid Sulfur: Conclusion

We conclude that liquid sulfur is not a promising candidate solvent for life. Evidence is very limited, but what evidence there is suggests that it will perform poorly on the criteria of Occurrence and Solute Stability, and performance on Solvation is essentially unknown.

## 3.5 Liquid Hydrogen Sulfide

Hydrogen sulfide ($H_2S$) has been suggested as a potential solvent for life in (Firsoff 1963, National Research Council 2019). $H_2S$ is the fourth most common gas emitted by terrestrial volcanoes (See SI, Section 1) and potentially on Io (Nash and Howell 1989, Russell and Kivelson 2001). We explore the potential of liquid hydrogen sulfide as a solvent for life in terms of Occurrence and other three categories in detail below.

### 3.5.1 Liquid Hydrogen Sulfide: Occurrence

Hydrogen sulfide is the fourth most common gas emitted by terrestrial volcanoes (after $H_2O$, $CO_2$ and $SO_2$). The fraction of sulfur outgassed by terrestrial volcanoes as $H_2S$ depends on the redox state of the volcanic vent system, and can be as high as 95%, although $SO_2$-rich, $H_2S$-poor volcanic outgassing predominates in Earth's oxidized crust (see SI Section 1). On a planet with a more reduced crust, $H_2S$ would be the dominant outgassed sulfur species. Despite its potential abundance on rocky planets $H_2S$ is a reactive chemical unlikely to accumulate as a liquid on the surface of a planet. We discuss these limitations on occurrence of hydrogen sulfide in detail below.

Unlike ammonia, hydrogen sulfide is not miscible with water (Selleck et al. 1952), it is poorly soluble in water (Chapoy et al. 2005) and water is relatively insoluble in liquid hydrogen sulfide (Santoli et al. 1999). It is therefore plausible that liquid hydrogen sulfide could exist on the surface of a rocky body where water is present (as ice), if the temperature and pressure conditions allow. However liquid $H_2S$ and liquid $CO_2$ are miscible (Francis 1954), and $CO_2$ is likely to be much more abundant that $H_2S$, even in the atmosphere of a planet with a reduced crust or mantle. Liquid $H_2S$ is therefore only likely to be the dominant solvent component in an environment where $CO_2$ is much rarer than $H_2S$, either because sulfur was much more abundant than carbon (which given the cosmic abundance of both elements is unlikely) or because carbon was overwhelmingly present as methane; hydrogen sulfide is readily soluble in liquid methane, and methane is readily soluble in liquid hydrogen sulfide, but the two liquids are immiscible (Kohn and Kurata 1958). Methane-dominated environment would require a Titan-like planet with a surface temperature above -60°C – a planet that might be termed a 'Cool Titan' by analogy with the reducing 'Warm Titan' atmosphere postulated for a post-impact early Earth (Zahnle et al. 2020). We note however that the terrestrial 'Warm Titan' model converts essentially all its $CH_4$ to $CO_2$ on a timescale of 2 – 20 MY (Zahnle et al. 2020) through atmospheric photochemistry.

More critically, $H_2S$ is photochemically short-lived in almost any anoxic atmosphere, whether dominated by $N_2$, $H_2$ or $CO_2$ (Hu et al. 2013). The vapour pressure over $H_2S$ even at its freezing point of -86°C is 0.21 bar (by comparison the vapour pressure of water at 0°C is 0.005 bar), so if an $H_2S$ ocean or lake is not to rapidly 'dry out' then the atmosphere above it must have at least 0.21 bar of $H_2S$. The required atmospheric pressure of 0.21 bar of $H_2S$ implies a high concentration in of $H_2S$ in the atmosphere to be photochemically destroyed. The photolysis rate of $H_2S$ is 100 times that of water (Hu et al. 2013). The products of the photolysis of $H_2S$ are hydrogen and elemental sulfur, they do not recombine to re-form $H_2S$.

As is it is in the case of ammonia discussed above, the photochemical reactivity of $H_2S$ is mirrored by its thermal lability (Gautam et al. 2024). Processing of the $H_2S$ reservoirs through volcanoes would result in removal of $H_2S$ by breakdown to $H_2$ and elemental sulfur.

Similarly to ammonia, the accumulation of hydrogen sulfide would require a minimal volcanic activity and a UV-shielded environment to prevent rapid photolysis. Such an environment would also shield methane from photolysis, so a UV-shielded, highly reduced 'Cool Titan' world could in principle accumulate stable liquid $H_2S$ lakes or oceans. The 'Cool Titan' surface rocks would also have to be largely devoid of transition metals or transition metal oxides, as $H_2S$ reacts readily with these chemicals to form sulfides or elemental sulfur (Poulton et al. 2002, Georgiadis et al. 2020). While such a planetary environment can be imagined, no precedent for it is known, and so we must consider the occurrence of liquid $H_2S$ as very unlikely.

There are no reported measures of the viscosity of liquid $H_2S$, but as its density, boiling point, dipole moment and hydrogen bonding potential are intermediate between those of methane and ammonia, we would expect a viscosity intermediate between these two solvents, and hence at the lower end of the range known to be compatible with life. We note however that to confirm this hypothesis the viscosity of $H_2S$ would need to be tested experimentally.

### 3.5.2 Liquid Hydrogen Sulfide: Solvation

Many organic and inorganic compounds stably dissolve in liquid hydrogen sulfide. These include alkanes, aromatic hydrocarbons, such as anthracene and naphthalene, alcohols, thiols and thio-ethers and trichloracetic acid (Quam 1925). Other compounds are stable to reaction with liquid $H_2S$ but insoluble. Examples of such insoluble compounds include citric acid, several sugars, cystine and glycine. However (Quam 1925) gives no quantitation of solubility. Most inorganic compounds appear to be insoluble in liquid $H_2S$ (Wilkinson 1931). For example, elemental sulfur itself dissolves only sparingly in liquid hydrogen sulfide (Smith et al. 1970). Whether soluble and insoluble moieties can be combined to form $H_2S$-amphipaths, a crucial characteristics of any biochemistry, has never been explored.

### 3.5.3 Liquid Hydrogen Sulfide: Solute Stability

As mentioned above, (Quam 1925) reports that many organic and inorganic compounds that are soluble in $H_2S$ are also stable. However their stability is not quantified. A number of compounds, both soluble and insoluble in liquid $H_2S$, react with it. Many of the tested compounds are also reactive with water, such as metalloid halides, $PCl_3$ and $SeCl_4$, although $SiCl_4$, which reacts almost instantly with water, is stably soluble in liquid hydrogen sulfide (Wilkinson 1931).

We are not aware of any studies of amphipathic or detergent properties of substances in liquid $H_2S$, but both the solvation properties listed above and the physical properties of liquid $H_2S$ suggest that liquid $H_2S$ amphipaths should exist. Liquid $H_2S$ does not show the dense, strong hydrogen bond network seen in water, but $H_2S$ molecules in liquid are nevertheless weakly orientated with respect to each other (Santoli et al. 1999), which will provide a weak entropic drive to inserting solutes into the solvent structure analogous to the hydrophobic interaction in water. The specific heat capacity of liquid $H_2S$ is similar to that of water (52 – 63 J/mol/K, depending on temperature, vs 75.6 J/Mol/K for water at STP), which also supports significant internal structure in liquid $H_2S$.

In summary, the limited experimental data suggests that diverse compounds will be stable when dissolved in liquid hydrogen sulfide.

### 3.5.4 Liquid Hydrogen Sulfide: Solvent Chemical Functionality

Liquid $H_2S$ can participate in a wide range of chemistry at Earth surface pressures and temperatures of -60$^{o}$C (e.g. (Quam 1925, Wilkinson 1931)). $H_2S$ is expected to be a much better medium for redox reactions than water, both because sulfur is a very redox-active element and because $H_2S$ can readily, and reversibly, form chains in liquid $H_2S$. Such polymerization reactions allow for a variety of chemical reactions that are unique to this solvent, including a polythiol-mediated redox and radical chemistry (e.g. (Dogadkin 1958, Huang et al. 2020)). Liquid hydrogen sulfide self-ionizes to form $H_3S^+$ and $HS^-$ ions to a much smaller extent than water or ammonia (Satwalekar et al. 1930, Wilkinson 1931), but can dissolve ionic compounds to provide for ion gradients; some organic thioacids such as dithioacetic acid dithiobenzoic acid and some base analogues such as triethylammonium hydrosulfide ionize in liquid hydrogen sulfide (Wilkinson 1931, Smith 1951). Such ionization implies that chemistry that relies on proton gradients or proton currents could be compatible with liquid hydrogen sulfide, adding to its potential as an active participant of biochemistry.

#### 3.5.5 Liquid Hydrogen Sulfide: Conclusion

We conclude that, like ammonia, liquid hydrogen sulfide has chemical properties, e.g. in terms of solvation, that make it a potentially promising solvent for life. However, $H_2S$ is very unlikely to occur as a liquid in any stable body on the surface of a rocky planet.

### 3.6 Liquid Hydrogen Fluoride

Liquid hydrogen fluoride (HF) has been proposed as a potential alternative solvent for life on planetary bodies that are substantially colder than Earth (Budisa et al. 2014). (National Research Council 2019) also lists hydrogen fluoride (HF) as a potential solvent for life. However, we consider HF to be implausible as a solvent for life for several reasons that we discuss in detail below.

#### 3.6.1 Liquid Hydrogen Fluoride: Occurrence

It is hard to imagine an environment where substantial amounts of HF would be present as liquid for geological periods of time, for two reasons. Firstly, while volcanoes outgas HF, they outgas far more water (See SI, Section S1). Therefore, for an HF ocean to form, as opposed to a water ocean containing HF, volcanic gases would have to have negligible oxygen while at the same time having at least some hydrogen. If they were completely depleted of water then there would be no hydrogen to form HF, but if hydrogen was present in the presence of oxygen water would form. It is not clear if an oxygen-depleted, hydrogen-containing planetary crust is realistic; the only possible scenario would be an extreme carbide planet, where carbon is more abundant than oxygen and oxygen is correspondingly outgassed exclusively as $CO_2$. Such planets have been postulated (Kuchner and Seager 2005), but whether they exist is unknown.

The second reason is that HF reacts with silica, so unless the putative HF-ocean planet had a crust composed of non-silicate, non-metallic material, any HF ocean would react with crustal rocks to form fluorides and water. On a carbide planet silica would be replaced largely by silicon carbide, which is also attacked slowly by HF (Habuka and Otsuka 1998). This reaction may be very slow at low temperatures, so a cold planet with a surface temperature near the melting point of HF (-83°C) with an excess of HF and no water could in principle accumulate HF lakes or oceans.

#### 3.6.2 Liquid Hydrogen Fluoride: Solvation

HF has been used experimentally as a solvent and catalyst in organic chemistry, and proteins and amino acids are surprisingly stable in liquid HF below 0°C (Lenard 1969, Norell 1970, Polazzi et al. 1974). HF is likely to dissolve metals as well. We are not aware of any systematic study of solvation of organics in hydrogen fluoride. (Gore 1869) provides a long list of inorganic materials' interaction with HF (over half react violently, most of the rest do not dissolve), and a sample of organics which stably dissolve including mono- and di-saccharides, caffeine, indigo and nitrocellulose. (Gore 1869) reports that moss and sponge was little affected by liquid HF, although the paper did not report if they actually survived as living material. Hydrogen fluoride forms very strong hydrogen bonds which form chain structures in liquid, rather than branching networks like the hydrogen bonds in water, sulfuric acid and ammonia (Maybury et al. 2004, McLain et al. 2004). However the hydrogen bonding structure of liquid HF is sufficient to drive a solphophobic effect, and micelle formation has been observed experimentally in liquid hydrogen fluoride (Zeiseler et al. 1992, Peters and Miethchen 1993, Roth et al. 1995).

#### 3.6.3 Liquid Hydrogen Fluoride: Solute Stability

As noted above, HF can stably dissolve some substances, in part because it remains liquid to -83°C, at which temperatures all chemistry will be very slow. (We note that the solubility of substances in liquid HF at these low temperatures has not been explored). The functional diversity of stable solutes in HF is not known.

#### 3.6.4 Liquid Hydrogen Fluoride: Solvent Chemical Functionality

HF is a protonating solvent but does not self-ionize. HF is a powerful oxidizing agent, and hence chemically active. HF may therefore meet some of the Chemical Functionality criterion.

#### 3.6.5 Liquid Hydrogen Fluoride: Conclusion

We conclude that hydrogen fluoride is an implausible solvent for life. While it passes on the criterion of Solvation and can plausibly be argued to meet Chemical Functionality, it is questionable on the criteria of Solute Stability and fails on Occurrence.

### 3.7 Formamide

Formamide has been discussed as both a reagent and a solvent for life, especially in the context of the origin of life (OOL) (Schoffstall and Laing 1985, Benner et al. 2004, Ziegler and Benner 2018, Gull et al. 2023). We conclude that formamide is chemically attractive as a solvent for life, but fails on the Occurrence criterion as it is unlikely to occur naturally.

#### 3.7.1 Formamide: Occurrence

Formation of pure formamide would have required a very specific set of circumstances to form at all. A scenario suggested by (Bada et al. 2016) requires the formation of a concentrated solution of formamide, cooling to between 0°C and 2.6°C to precipitate out pure formamide, removal of the water and then re-warming to form liquid formamide. Other scenarios that lead to the formation of abundant formamide are likely to be equally convoluted. Thus, while liquid formamide might have occurred in rare, specific circumstances, it is extremely unlikely to provide for long-term lakes or oceans of formamide as a solvent in the planetary context. Formamide is also photochemically labile (Boden and Back 1970), breaking down to CO, $H_2$ and $NH_3$, the $NH_3$ itself being photochemically labile. Thus, even if formamide lakes did form, they would not be stable to long-term photochemical destruction of the formamide from stellar UV.

### 3.7.2 Formamide: Solvation

Formamide is a powerful solvent, dissolving a wide range of materials. It has a high dielectric constant which allows it to effectively dissolve polar molecules, but it is not hydrogen bonding and so can dissolve more hydrophobic molecules than can water. While it dissolves a larger breath of molecules than water, it does not dissolve everything across the board. Some amphipaths do form micelles in formamide (e.g. (Couper et al. 1975, Akhter and Alawi 2000)), and some substances, for example, the complex hydrocarbons in pitch, are virtually insoluble in formamide (Papole et al. 2010) as are some polymers (e.g. (Jousset et al. 1998)). Many metal salts are soluble in formamide, although the patterns of solubility differ from water; for example, ammonium chloride is very soluble in water but practically insoluble in formamide (Magill 1934). Formamide therefore fulfills the solvation criteria of being able to dissolve a range of compounds, but not all compounds.

### 3.7.3 Formamide: Solute Stability and Solvent Chemical Functionality

Formamide is relatively unreactive, and is not seen to react with a wide range of solutes. It is a participant in some chemistry, as noted above in the OOL literature. Materials such as hexamethyldisilazine, which would be rapidly hydrolysed in water, are stable (although insoluble) in the presence of formamide (Grunert 2002). Formamide therefore can stably dissolve a larger chemical space of molecules than water, but it itself provides only limited chemical functionality.

### 3.7.4 Formamide: Conclusion

We conclude that formamide is an unlikely solvent for life. It can provide for Solvation, Solute Stability and some Chemical Functionality, but is extremely unlikely to occur on planetary surfaces.

## 3.8 Carbon dioxide

Carbon dioxide as a potential solvent for biochemistry is considered in two phases, as a liquid and as a supercritical fluid.

### 3.8.1 Carbon Dioxide: Occurrence

$CO_2$ is a common and widely distributed compound, being the dominant component of the atmospheres of Mars and Venus, a significant component of the Earth's atmosphere and the second most common volatile emitted from terrestrial volcanoes (See SI, Section S1). $CO_2$ is liquid over a range of temperatures and pressures, although not a range that is found on Earth's surface today. Liquid $CO_2$ can be found in Earth's ocean floor sediments at 4 $^oC$ and tens of bar pressure, resulting from accumulation of volcanic $CO_2$ (Sakai et al. 1990, Konno et al. 2006, de Beer et al. 2013). Submarine liquid $CO_2$ is quite a rare find, as it is less dense than water, and so it tends to leak upwards as 'bubbles' to higher in the ocean, where the lower pressure allows it to boil into $CO_2$ gas. It has been suggested that the cold, dense atmosphere of early Mars could have resulted in liquid $CO_2$ flow, and that liquid $CO_2$ could form at limited sites in the recent past (Hecht et al. 2024). (Graham et al. 2022) studied whether a 'Cold Venus' planet, with a dense $CO_2$ atmosphere but a low surface temperature, could have a liquid $CO_2$ ocean under its $CO_2$ atmosphere. Unlike the Earth, where the boiling point of the constituents of the ocean and the atmosphere are quite distinct, the Cold Venus' ocean would expand or contract substantially with changes in insolation, pressure and 'weather'. However (Graham et al. 2022) considered that the ocean could be stable over geological time. The 'Cold Venus' would form through the same scenario envisaged for a desiccated red dwarf planet above (Section 3.3.1), but with a more distant orbit and without the requirement to remove all the planet's water, as the solubility of water in $CO_2$ is quite low.

The other phase in which $CO_2$ has been discussed as a solvent for life is as a supercritical fluid (Budisa and Schulze-Makuch 2014). Supercritical $CO_2$ occurs in the atmosphere of Venus in our own Solar System, but we should be aware that Venus' ground-level atmosphere is not a solvent for life, for two reasons. The first, an obvious reason, is that it is too hot to allow the stable existence of organic compounds. The second is that 'supercritical' simply means above the critical temperature and pressure. There is no phase change in such fluids on compression, so a supercritical fluid under sufficient pressure can have a high density without being a liquid. The term 'supercritical fluid' is often taken to mean substances that have a density that approaches that of the liquid phase because they are near their critical temperature and have been compressed to high pressure. Supercritical $CO_2$ in this sense has been suggested as a solvent for life (Budisa and Schulze-Makuch 2014). Venus' ground-level atmosphere, while technically supercritical, does not meet this criterion. Supercritical $CO_2$ needs to be *dense* supercritical $CO_2$ to dissolve large molecules, especially polymers. Such scenario is discussed further below.

The link between density, temperature, pressure and solubility of substances is expanded on in the SI, Section S4. Here we note that to be a plausible solvent for life supercritical $CO_2$ must be under at least 100 bar pressure *and* be at moderate temperatures, probably below 100$^o$C, but above the critical temperature of 30.98$^o$C. The powerful greenhouse effect of $CO_2$ makes this a difficult scenario to fulfil in any realistic planetary environment.

#### 3.8.2 Carbon Dioxide: Solvation

$CO_2$ has no permanent dipole and low polarizability and does not form hydrogen bonds, and so its solvent behaviour is similar to organic solvents such as toluene, dissolving hydrophobic, non-polar molecules (Hyatt 1984). Liquid $CO_2$ is known to be a good solvent of a wide range of substances, including hydrocarbons, esters, crown ethers, substituted benzenes and pyridines, but not sugars or metal salts (Francis 1954, Gouw 1969, Hyatt 1984). Chelated metals including Co, Cu, Fe, Mn and Zn are known to be soluble in supercritical $CO_2$ (Yazdi and Beckman 1994, Ashraf-Khorassani et al. 1997, Smart et al. 1997). Some materials are amphipathic in liquid $CO_2$ (van Roosmalen et al. 2004), and can be used as detergents to solubilize polar molecules into liquid $CO_2$ (Cooper et al. 1997).

Supercritical $CO_2$ is widely used in industry as an extraction solvent, and solubility of over 1600 substances including ionic solids in supercritical $CO_2$ has been measured (Gupta and Shim 2006).

Liquid $CO_2$ has a density of >~800 kg/m$^3$ (depending on temperature and pressure – see SI, Section 4). The minimum density needed for $CO_2$ to effectively solubilize molecules of the sizes typically found in biological materials has been explored in the industrial use of liquid and supercritical $CO_2$ as a dry cleaning fluid and as an extractant (e.g. (Banerjee et al. 2012)). Densities of at least 400 kg/m$^3$ (half that of liquid $CO_2$ at 1 bar pressure, 40% that of water) are generally required, and >800 kg/m$^3$ for solubilization of polymers such as proteins. Liquid $CO_2$ fulfils these criteria, but supercritical $CO_2$ only fulfils them at high pressures at temperatures below 200°C; for example, at the average surface temperature of Venus (460°C) the density of supercritical $CO_2$ only exceeds 400 kg/m$^3$ at ~660 bar (7 times the surface pressure of Venus), and exceeds 800 kg/m$^3$ at ~1500 bar. As noted above, only a very specific set of planetary environments would allow these conditions to occur.

In addition, the ability of near-critical $CO_2$ to dissolve substances changes dramatically with changes in pressure (unlike liquid $CO_2$, or liquid water), as illustrated in SI Section 4. Dramatic changes in solvation with relatively small changes in temperature would make it effectively impossible to build structures that depended on differential solubility of their components, such as lipid bilayers or globular proteins, in an environment where the pressure or temperature can change.

#### 3.8.3 Carbon Dioxide: Solute Stability

A wide range of materials can be stably dissolved in liquid and supercritical $CO_2$. These include chemically sensitive materials such as proteins that are likely to be damaged by other extractive or processing methods (Quirk et al. 2004, Woods et al. 2004). Supercritical $CO_2$ can be a medium in which biological materials can be hydrolysed by water at elevated temperatures (e.g. (Brunner 2009, Yakaboylu et al. 2015, Zhao et al. 2019)), but at low temperatures or in the absence of corrosive substances like water $CO_2$ is a very chemically benign solvent.

#### 3.8.4 Carbon Dioxide: Solvent Chemical Functionality

$CO_2$ as a solvent is unlikely to participate in the chemistry of life. $CO_2$ is quite chemically inert, and does not self-ionize. Therefore, it is unlikely to participate in biochemistry, beyond its role to solvate life's component molecules.

#### 3.8.5 Carbon Dioxide: Conclusion

We conclude that liquid carbon dioxide is a potential solvent for life if a planetary environment can be found in which it occurs stably as a liquid. It fulfils the criteria of Solvation and Solute stability, although it fails on Solvent chemical functionality. By contrast supercritical $CO_2$ appears less plausible as a solvent, due to the quite restricted range of conditions under which it could solubilize complex molecules and have stable, unchanging solvation properties.

## 3.9 Liquid Sulfur Dioxide

Sulfur dioxide has not hitherto been considered as a solvent for life, although liquid $SO_2$ has been considered as a solvent in one origin of life scenarios (Sydow et al. 2017, Sydow et al. 2023).

#### 3.9.1 Sulfur Dioxide: Occurrence

$SO_2$ has a critical temperature of 157°C, and could form a non-protonating polar liquid under that temperature with sufficient pressure (Burow 2012). On Earth $SO_2$ is the third most common gas in volcanic outgassing (See SI, Section S1), and may be the second most common volcanic gas on Venus. A 'Cold Venus' scenario as outlined above could therefore result in sufficient atmospheric pressure and $SO_2$ abundance to accumulate liquid $SO_2$.

However, such a planet would have to be even more thoroughly dehydrated than the case suggested for sulfuric acid above. Even trace amounts of water would result in $SO_2$ forming sulfurous acid which would disproportionate to sulfuric acid. Photochemical oxidation of $SO_2$ to $SO_3$ would also result in sulfuric acid if there was even a trace of water in the atmosphere; on Venus photooxidized $SO_2$ forms sulfuric acid even though the mesosphere has only less than 10 ppm water (Bains et al. 2021a). It is not clear if a complete stripping of all water (which means all hydrogen atoms) from the atmosphere and crust of a planet is possible.

#### 3.9.2 Sulfur Dioxide: Solvation

Liquid $SO_2$ has been widely studied as an industrial solvent, and is known to dissolve a range of organic compounds and inorganic ions, including dimethylsulfide, benzene, naphthalene, and carbon tetrachloride. However propane and butane are almost insoluble in liquid $SO_2$, suggesting that amphipaths in liquid $SO_2$ may exist (Ross et al. 1942, Burow 2012).

#### 3.9.3 Sulfur Dioxide: Solute Stability and Solvent Chemical Functionality

The chemical functionality of liquid $SO_2$ is expected to lie between those of water and $CO_2$, in that $SO_2$ can engage in redox chemistry and addition reactions to carbonyl groups, but it is not a protonating solvent. $SO_2$'s ability to complex carbonyls is the reason for its consideration as a key player in some Origin of Life scenarios (Sydow et al. 2023). $SO_2$ can also form solvent complexes with thioethers such as dimethylsulfide, and with amines and ethers. This could point to liquid $SO_2$ having quite rich chemistry aside from its role as a solvent.

#### 3.9.4 Sulfur Dioxide: Conclusion

We conclude that sulfur dioxide is not attractive as a candidate solvent for life. While it passes the Solubility and Solute stability criteria, and has promise for the Solvent Chemical Functionality criterion, it is very likely to fail the Occurrence criterion as a pure solvent in its own right, in any realistic planetary scenario.

### 3.10 Cryogenic Solvents (Methane, Ethane, Nitrogen)

We discuss substances that are only liquid below -80°C together as 'cryosolvents', as they share similar characteristics. The cryosolvents include methane, ethane and their admixture on the surface of Titan (Hayes 2016), and liquid nitrogen postulated to form under nitrogen ice on Triton (Soderblom et al. 1990). All have been suggested as solvents for life (Bains 2004, McKay and Smith 2005, Norman 2011, McLendon et al. 2015, Stevenson et al. 2015, McKay 2016). However, their almost total inability to actually act as a solvent to anything but the smallest molecules makes them very unlikely candidates for solvents for life.

Can the chemistry of life function in solvents substantially below the freezing point of water? This question applies to most of the solvents considered so far, but it is especially relevant to the cryogenic solvents. Reaction rates decline exponentially with temperature, with typical metabolic processes decreasing 2 – 4 fold with a 10°C drop in temperature (the so-called $Q_{10}$ value, e.g. (Davidson et al. 2006, Schipper et al. 2014) and references therein). A $Q_{10}$ of 3 with 10°C would mean reactions occurring at ~$5 \cdot 10^{-9}$ times their terrestrial rate in liquid nitrogen, which is consistent with liquid nitrogen's use for stable storage of labile biological materials in the laboratory. However, this is the comparative rate of the same reactions between the two temperature regimes. Terrestrial life adapts to temperatures between -15°C and +122°C, with fairly similar growth rates at both temperatures (Clarke 2014). Life adapts its growth rate to both temperature extremes by changing the chemistry of the cell, primarily through adaptation of the catalysts but also in some cases the metabolites. If we accept that life in a non-aqueous solvent would have chemistry different from that of the Earth, then we must also accept that the chemistry of that life would be adapted to function at whatever rate is evolutionarily selected in that solvent. These might include utilization of chemicals that are highly unstable under terrestrial temperature regimes but which are only moderately reactive under cryogenic conditions, i.e. which reaction rates and stability is 'tailored' to the cryogenic conditions. Examples of such compounds include methanimine (stable at 20 K) (Guillemin and Denis 1988), tetrazane (stable at 70 K) (Rice and Scherber 1955), triphosphane (stable at 200 K) (Baudler 1982) and diamino triphosphane (stable in liquid ammonia) (Baudler 1987). We do not suggest a metabolism based on these compounds, but rather use these examples to illustrate that chemicals that react at moderate speed at very cold temperatures are known, and therefore that the extrapolation of terrestrial biochemical reaction rates to cryogenic temperatures is not a thorough exploration of the chemistry that could occur at those temperatures.

### 3.10.1 Cryogens: Occurrence

(Ballesteros et al. 2019) predict that liquid ethane should be more abundant in planetary environments than any other liquid, and liquid ethane, methane and nitrogen between them should have 13 times the abundance of liquid water on the surface of planets and moons, including those around diverse star types. Ethane and methane are subject to photochemical destruction, and so will need a constant source or a regeneration mechanism to maintain them (Coustenis 2021). Nitrogen is extremely stable, as noted above, but has a critical temperature of -145°C, so for a planet to maintain significant surface nitrogen as a liquid (unlike Triton's transient $N_2$ geysers) it would need an atmosphere that remains gaseous at below -145°C, i.e. one composed of hydrogen (or neon or helium). In addition, the planet would have to be sufficiently far from its star, and the internal radiogenic heat budget is sufficiently small, that the surface remained below nitrogen's critical temperature despite hydrogen's substantial greenhouse effect. No-one has explored whether such a scenario is plausible.

### 3.10.2 Cryogens: Solvation

All the cryogenic solvents are both non-protonating and apolar. They are also, by definition, very cold. As solubility of solids in liquids declines with temperature, it is unsurprising that very few molecules are soluble in cryogens. Thus for example liquid methane, despite being sometimes considered as a possible solvent for life, is unlikely as a solvent because very few molecules dissolve in it (Petkowski et al. 2020). No polymer systems have been found that are soluble in cryogenic liquids (McLendon et al. 2015). Diverse small molecules and polymers can dissolve in higher hydrocarbons (McLendon et al. 2015), such as butane, which has a similar boiling point to ammonia, or in octane. However, we do not consider higher hydrocarbons such as octane further here, as we are aware of no scenario which would provide a planet with stable liquid octane reservoirs in the absence of life.

In principle biochemistry in cryogens could use a much wider range of chemical bonds than chemistry in water. The reactive chemistry of water substantially limits the chemical bonds that life can use; in cryogens the low temperature would render dissolved water much less reactive, and water itself would be expected to be poorly soluble (although experimental measurements suggest that water is more soluble than expected in cryogens (Rebiai et al. 1984)). Thus, exotic chemistry including silicon, germanium and selenium bonded to each other and to nitrogen, sulfur and halogens are expected to be stable in cryogens, greatly expanding the chemical space available to life in cryogens. However it is doubtful whether the larger chemical space available in cryogens can compensate for the much lower solubility of all but the smallest molecules (Petkowski et al. 2020). Early work suggesting that bilayer type multimolecular structures can be formed in cryogens from small nitrogen-containing molecules (Stevenson et al. 2015) has since been disputed (Sandström and Rahm 2020).

#### 3.10.3 Cryogens: Solute Stability and Solvent Chemical Functionality

At cryogenic temperatures ethane, methane and nitrogen are completely inert. To the limited extent that they can act as solvents, they are purely support for the molecules they dissolve and will not participate in those molecules' chemistry.

#### 3.10.4 Cryogens: Conclusion

We conclude that the cryogenic solvents methane, ethane and nitrogen are very unlikely to be solvents for life. While they may be common, they fail on Solvation, as well as on Chemical Functionality criteria.

## 3.11 Other Suggested Solvents

Several other liquids have been suggested as candidate solvents for life, which we will discuss briefly here. Hydrazine (Schulze-Makuch and Irwin 2004) is thermodynamically unstable to decomposition to nitrogen and hydrogen, more photolabile and substantially more reactive than ammonia, and so is unlikely to occur in substantial amounts on rocky planets. (Tang et al. 2006) have suggested that lakes or oceans of methanol could have existed on early Mars, based on similarities between Martian hematite and $FeCl_3$ treated at low temperatures with methanol. The methanol is proposed to come from photochemical oxidation of methane. However the photochemical destruction of methanol is much faster than its synthesis from methane (See SI, Section 6), especially at the low temperatures postulated by (Tang et al. 2006), so accumulation of methanol in any rocky planet environment seems unlikely (Huang et al. 2022b). Hydrogen cyanide has been suggested as a solvent, and can be generated in nitrogen-rich atmospheres in atmospheres with a high carbon to oxygen ratio (Rimmer and Rugheimer 2019). HCN is a key precursor molecule in a number of scenarios for the origin of life (Sasselov et al. 2020). However, lakes or oceans of relatively pure HCN seem unlikely. Liquid hydrogen cyanide is liable to runaway polymerization to a mixture of oligomers, some of which are stable (such as adenine) but some of which are themselves liable to explosive decomposition (Williams and Schwartz 2014). It is implausible that an ocean of liquid HCN could survive explosive polymerization triggered by lightning strikes, volcanic activity or meteorite impact. Oxygen difluoride ($F_2O$) has been suggested because of its similar shape and dipole to $H_2O$ (Firsoff 1963, Forward 2012). However, $F_2O$ cannot act as a planetary solvent for many reasons. It reacts explosively with water, will displace oxygen from oxides and chloride from chlorides, and it decomposes at 200°C to elemental fluorine and oxygen, so even if a mechanism could be found to generate it outside human industry its survival as a bulk liquid seems very unlikely.

### 3.12 Summary of Survey of Solvents

Our conclusions are summarised in Table 1. We conclude that water is indeed the most plausible solvent for life, but concentrated sulfuric acid is also a realistic alternative. Liquid $CO_2$ is an interesting possibility, if sufficiently diverse chemicals can be found that will dissolve in it, and if its lack of chemical functionality can be overcome (or proven illusory). Liquid sulfur remains an outside possibility, as no plausible scenario for an environment where it is stably present as a mobile liquid is known, and there is little known about solvation or solute stability. Ammonia, hydrogen sulfide, formamide, hydrogen fluoride, supercritical carbon dioxide, sulfur dioxide and the cryogenic solvents appear to be ruled out as realistic candidates for solvents for life by this analysis.

The ranges of temperatures over which these materials are liquid is summarized in Figure 1. The freezing point of liquids is not materially affected by pressures relevant to the surface of rocky planets (in the order of zero to 1000 bar), but the boiling point changes substantially with pressure. Figure 1 therefore shows the temperature at which each liquid boils for several different pressures. The upper limit is the critical temperature, above which a substance will not form a distinct liquid phase no matter how much the pressure is increased.

In general substances with lower freezing and boiling points have smaller liquid ranges, but this need not limit their relevance as a solvent for life. Environments cold enough to support cold liquids also are likely to have smaller temperature ranges. For example, the Earth has surface temperatures from 220 K to 320 K, a range of 100 K, while Cassini measured the temperature range on Titan to be only 90 – 94 K, a range of 4K. (Jennings et al. 2019) (we also note that water is liquid for less than 50% of Earth's surface temperature range). The environmental temperature range is also heavily modulated by the planetary atmosphere. Venus, with a dense opaque atmosphere shows only a few K difference in surface temperature between day and night, equator and poles (Basilevsky and Head 2003). The conclusion from Figure 1 should not be that one solvent is more plausible than another because it has a larger range over which it is liquid. Rather, the range over which a solvent can be liquid is an initial indication of the range of planetary environments where that liquid might be found.

We note that the relevance of the specific heat capacity and consequent thermal buffering ability of a solvent, often cited as a strong argument for the unique properties of water (e.g. (Pohorille and Pratt 2012)), is also dependent on the atmosphere. Strong thermal buffering is not needed on Titan or Venus.

## *4 Solvent mixtures*

The discussion above primarily addresses solvents as if they were each a single substance. However chemical species rarely occur in nature in pure form, and this applies to solvents as well. In the case where two solvents mix or dissolve in each other, the ratio that determines which solvent's chemistry dominates depends on the solvents concerned. Thus, for example, 80% w/w sulfuric acid has chemistry similar to other acid solutions in water, whereas 95% w/w sulfuric acid has quite different chemistry.. Therefore, for the sulfuric acid/water system, the crossover between the solvent being water and the solvent being sulfuric acid lies somewhere between 80% w/w and 95% w/w. For liquid ammonia, anhydrous liquid ammonia can solvate alkali metals and electrons, but even 1% of water in liquid ammonia means that alkali metals react to form hydroxides and release hydrogen. Thus, for the ammonia/water system, the crossover between the solvent being water and the solvent being ammonia may lie at 99% ammonia.

Could mixtures of solvents provide substantially different chemistries from the pure solvents discussed above, and hence could they provide additional solvent options for life? We think this in unlikely, with two exceptions.

Firstly, in any plausible planetary scenario, one substance will dominate over another in terms of mass or abundance. In the case of ammonia, it is likely that a rocky planet will have more water than ammonia, both because water is more cosmically common and because water is more stable to photochemical breakdown.

Secondly, the chemical properties of one solvent will probably dominate the other, as the example of ammonia:water mixture discussed above illustrates.

There are two exceptions worth mentioning.

The first exception is freezing point depression of solvent mixtures. Any solute will tend to depress the freezing point of the liquid in which it dissolved. This is true of salt in seawater, and of solutions of ammonia in water or water in ammonia. Pure water freezes at 273 K, pure ammonia at 195 K, but water:ammonia mixtures can remain liquid down to 176 K (Chua et al. 2023). Similarly, a 50:50 mixture of hydrogen peroxide and water has a freezing point of -52°C (Foley and Giguère 1951), below the average surface temperature of Mars (Houtkooper and Schulze-Makuch 2007). Pure sulfuric acid freezes at 283 K, but 94% w/w sulfuric acid (the rest being water) freezes at 241 K (DKL Engineering Inc. 2024). Thus, the lower limit on relatively high temperature solvents such as sulfur, sulfuric acid and water could be reduced by admixture of either other solvents or other solutes. This could extend the planetary range over which they are considered.

The second exception is sulfuric acid. While the chemistry of <70% w/w sulfuric acid in water is predictable from the chemistry of other acid solutions in water, in the range 70% w/w to 100% w/w sulfuric acid's chemistry changes such that 90% w/w sulfuric acid is a distinct solvent unlike either 100% w/w acid or 70% w/w acid (Liler 2012). This unique behavior of sulfuric acid at different concentrations is because, unlike other pairs for which the chemistry is well known, water and sulfuric acid mutually react. The distinct characteristic of the $H_2SO_4$ solvent at different concentrations has significant implications for life's use of sulfuric acid as a solvent, as mentioned in section 3.3.2.

The example of sulfuric acid as a solvent also introduces the concept of "solvent replacement". One model for how life could have come to use sulfuric acid as a solvent on Venus is that it originated in the hypothetical water ocean during the early phase of Venus' history (Way et al. 2016) and then adapted to use sulfuric acid as a solvent instead of water when sulfuric acid became the dominant available liquid. We discuss the concept of solvent replacement in the next section.

## 5 Solvent replacement.

In the discussion above we have assumed tacitly, as do all other discussions, that for a planet to be habitable it must have a solvent on (or under) its surface or in its clouds, and that that solvent must be the same for the entire duration of planet's habitability. Thus Mars is considered to have become uninhabitable on its surface when it lost its surface water (Jakosky 2021), and Earth will similarly become uninhabitable when it loses its surface water due to increasing insolation (Caldeira and Kasting 1992).

However, the Earth may be atypical in having a surface environment in which just one solvent was abundant since its surface cooled. Venus may have had liquid water on its surface (Way et al. 2016), but now has sulfuric acid cloud droplets. Red dwarf planets may suffer dramatic loss of water to become 'cold Venuses' with liquid $CO_2$ oceans if planetary migration towards their host star does not offset the decline in instellation as the star cools. There may be a range of other environments in which it would be advantageous for native life to switch from using one solvent to using another solvent as the basic solvent for life. Could such solvent replacement happen?

For life to change its solvent, every aspect of its biochemistry must adapt to the new solvent. This might appear to be too extreme a change to be possible. However, two lines of evidence suggest otherwise.

Firstly, many components of terrestrial biochemistry can function in solvents other than water, despite billions of years' of adaptation to function in water. A wide range of enzymes can function in inorganic or even apolar solvents (Klibanov 1989, Volkin et al. 1991, Gupta 1992, Wescott and Klibanov 1994, Schmitke et al. 1996), albeit at drastically reduced efficiencies (as would be expected of catalysts optimized to work in water). DNA can form stable double helices in anhydrous glycerol (Bonner and Klibanov 2000) and in formamide, albeit with a lower melting temperature in the latter than in water (Casey and Davidson 1977, Blake and Delcourt 1996). Entire membrane-bound organelles can retain their structure in anhydrous glycerol (Siebert and Hannover 1978). Therefore, the flexibility is present in at least some biochemical systems to adapt to other, chemically similar solvents, even in the absence of selective pressure to do so.

Secondly, terrestrial life shows some adaptability to other solvent environment. The adaptation of yeast to grow in up to 25% ethanol is well known. In vitro evolution for ethanol tolerance can select for increased tolerance in 100 - 200 generations (Voordeckers et al. 2015, Mavrommati et al. 2023), suggesting that adaptation to a substantially altered solvent does not require whole-scale alteration of the biochemistry of the cell, but quite wide-spread 'rewiring' of metabolism does occur (Ma and Liu 2010, Snoek et al. 2016). Nor is adaptation achieved by excluding ethanol from the cell; on the contrary, some yeasts accumulate ethanol inside the cell at concentrations over that outside the cell (D'Amore and Stewart 1987), in some cases over 10-fold concentration (Legmann and Margalith 1986). As would be expected, then, some highly ethanol tolerant yeast grow poorly in solutions containing no ethanol; they have genuinely adapted to their new solvent environment (e.g. (Flor and Hayashida 1983, Jimenez and Oballe 1994)).

The example of yeast adaptation to ethanol directly addresses whether membrane bilayers could be adapted to a different solvent environment. Ethanol's toxic effects are primarily due to the chaotropic effects of ethanol on lipid bilayers (Ball and Hallsworth 2015), but yeasts can adapt their membranes to grow in 25% ethanol. No yeasts that can grow in 30% ethanol are known, but this limit may represents a limitation on the environments in which yeasts would naturally evolve rather than an absolute chemical limit.

Xerophilic fungi can also accumulate up to 25% glycerol inside their cells in response to water stress at 25°C (Hocking and Norton 1983, Hocking 1986, Hallsworth and Magan 1995) and high glycerol actually improves the fitness of some organisms growing at low temperatures (Chin et al. 2010). Glycerol is a polar hydrogen bonding solvent like water, but for that reason is considered a chaotropic agent that disrupts protein structure; despite this, a range of yeast and bacteria can adapt to up to 25% total wet weight of glycerol as an internal solvent.

A less well explored example is the adaptation of organisms to growth in heavy water. Replacement of hydrogens with deuterium can have substantial effects on metabolism (Navratil et al. 2018, Pirali et al. 2019), and consequently organisms usually grow less efficiently in heavy water than isotopically normal water. Many bacteria have been adapted to grow in heavy water (Mosin and Ignatov 2014), and yeast adapted to grow in heavy water show poorer growth in normal water (Kampmeyer et al. 2019), again showing genuine adaptation to a new solvent environment rather than increased tolerance for a range of solvent environments.

We emphasise that neither the precedent of terrestrial biochemical function in non-aqueous solvents nor the examples of terrestrial life adapting to slightly changed solvent environments provide examples of a change from one solvent to a completely different solvent. Yeast and bacterial adaptation of internal ethanol and glycerol has not been demonstrated beyond 25%, and heavy water is only very slightly chemically different from 'light' water. However, these examples show that at least some limited solvent change is possible, and complete solvent replacement, at least within the category of polar solvents, is at least worth considering and investigating. Exploring more extreme solvent changes (for example, growing organisms in increasing ethanol concentrations, or in the presence of dimethyl sulfoxide (DMSO) or dimethylformamide (DMF)) would be a valuable 'in vitro evolution' experiment of the sort well established in microbial evolutionary genetics (Hindré et al. 2012, Good et al. 2017, Lenski 2017)

Two astrobiological speculations may illustrate the relevance of solvent replacement to adapt to environmental change.

(Houtkooper and Schulze-Makuch 2007) suggested that Martian organisms might make hydrogen peroxide ($H_2O_2$) to form an $H_2O_2/H_2O$ solvent mix, and that the presence of organisms contain g substantial $H_2O_2$ could explain some aspects of the Viking Lander results. As mentioned above, the majority of the water in an actively, aerobically metabolising E.coli cell is derived from its internal metabolism, so it is not implausible that an organism could instead generate a substantial concentration of $H_2O_2$ as a solvent. Such a mixture has a lower freezing point than water, and $H_2O_2$ is mildly hygroscopic and so could help to abstract water from the environment. This is an intriguing idea, but an ecosystem that makes its own volatile solvent that is prone to loss is inherently unstable. Therefore, it is quite possible that a Martian organism (should such exist) makes $H_2O_2$ as a minor component of its internal liquid, but less likely that $H_2O_2$ would be a major component of Martian life's internal solvent.

An intriguing possibility for solvent replacement is the class of compounds called ionic liquids. Ionic liquids are salts with a melting point below 100$^{o}$C (Hajipour and Rafiee 2009, Lei et al. 2017), and all known examples are complex chemicals that are unlikely to occur naturally. They can have all the properties necessary to be a solvent for life that we discussed above except natural occurrence; ionic liquids are synthetic products, often with relatively complex chemical structures compared to naturally occurring liquids (See SI, Section S5)

Ionic liquids have extremely low vapour pressure, so a droplet of ionic liquid will never 'dry out' even under hard vacuum at Earth ambient temperatures (Horike et al. 2018). As a result, the risk of permanent desiccation for an organism using an ionic liquid as a solvent could be very small, leaving only the risk of mechanical damage as a cause of solvent loss to the outside world. An organism adapting to desiccation by accumulating charged osmolytes might end up with its internal solvent composed entirely of those osmolytes as an ionic liquid.

**Table 1**. Summary of solvents' match to the criteria discussed in this work. Green tick = meets that criterion. Amber tick – probably meets the criterion, but there is experimental or observational uncertainty. Red cross – does not meet that criterion. Amber cross – probably does not meet that criterion, but there is experimental or observational uncertainty.

| | Criteria | | | |
|---|---|---|---|---|
| | Occurrence | Solvation | Solute stability | Chemical functionality |
| | Is the solvent likely to stably occur on / near rocky planet (or moon) surfaces over geological time scale | Does the solvent dissolve a range of molecules (including ions), and not dissolve other molecules | Are chemically diverse chemicals stable when exposed to the solvent | Is the solvent capable of participating in metabolism (as opposed to just dissolve biochemicals) |
| Water | √ | √ | √ | √ |
| Ammonia | X | √ | √ | √ |
| Conc. Sulfuric acid (>90%) | √ | √ | (√) | √ |
| Liquid sulfur | (X) | (√) | (X) | (√) |
| Hydrogen fluoride | X | √ | (X) | √ |
| Formamide | X | √ | √ | √ |
| Liquid carbon dioxide | √ | (√) | √ | X |
| Supercritical carbon dioxide | X | √ | √ | X |
| Sulfur dioxide | X | √ | √ | (√) |
| Hydrogen sulfide | X | (√) | (√) | √ |
| Cryogenic solvents | √ | X | √ | X |

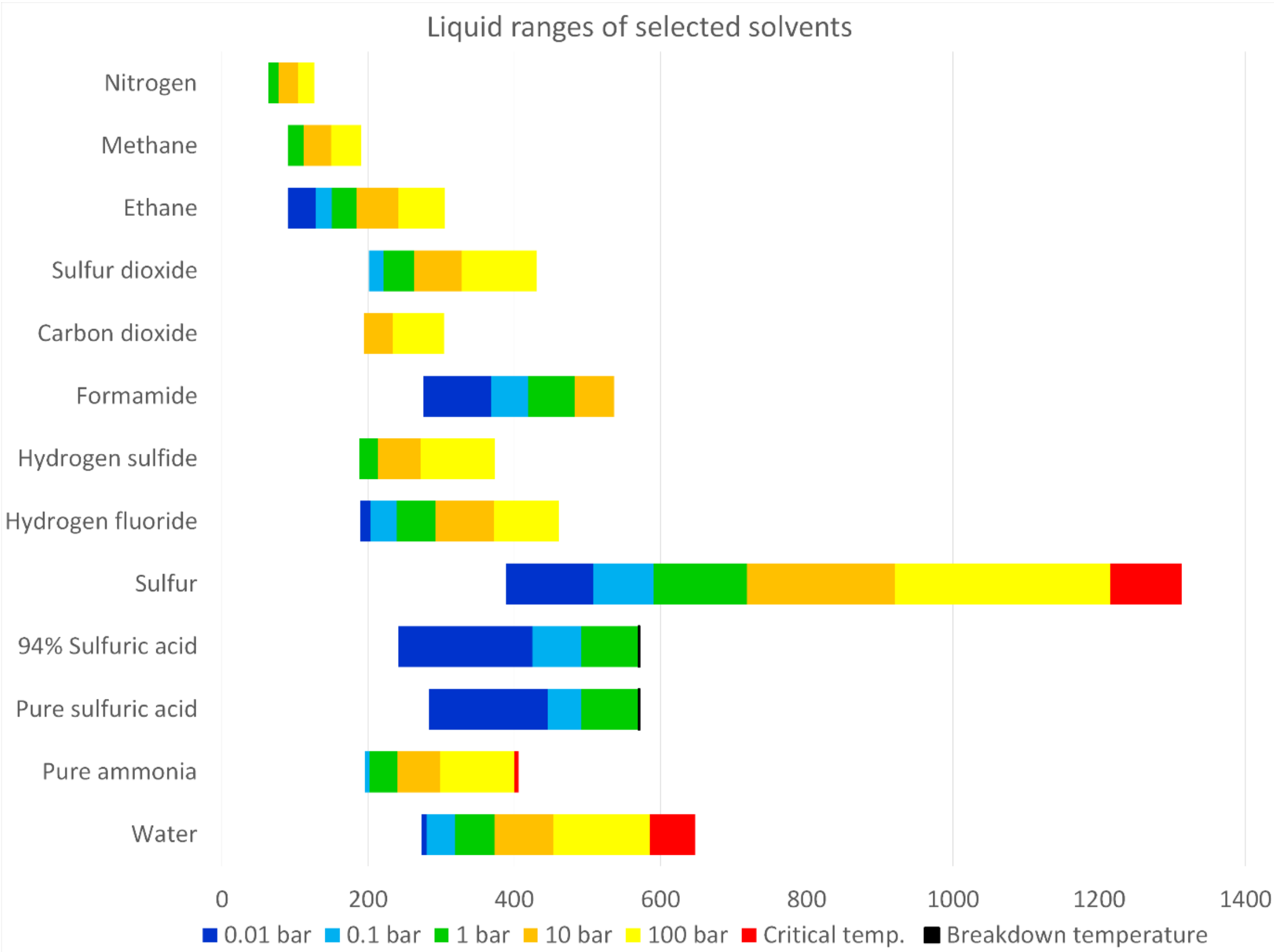

**Figure 1.** Liquid ranges for the solvent candidates. Y-axis: solvent candidate, X-axis: temperature (K). Data freezing points and Antoine equations for vapor pressure as a function of temperature from (Yaws 1999), supplemented by critical temperature data from (Compressed Gas Association 1990a, Compressed Gas Association 1990b). Boiling point for 94% sulfuric acid was calculated from data from (Ayers et al. 1980). For each substance the left end of the bar is its freezing point, and colored bars indicate the range at which it is liquid at 0.01 bar (dark blue bar), 0.1 bar (light blue bar) etc. Note that some substances such as carbon dioxide cannot be liquified at lower pressures; in the case of carbon dioxide the solid sublimes directly to a gas at Earth surface pressures, an so the left-most segment of its liquid range bar is the 10 bar color, orange. Only water, ammonia and sulfur can boil at temperatures above their boiling point at 100 bar. Sulfuric acid (here in w/w %) decomposes at ~570 K, and so cannot form a stable liquid above this temperature.

## *6 Conclusions.*

We have provided a framework for the assessment of potential solvents for life, and surveyed liquids and supercritical fluids that have been suggested as candidate solvents for life within that framework. Water remains the most likely solvent for life. Perhaps unexpectedly, concentrated sulfuric acid meets all the criteria for a solvent for life, although uncertainty remains about the diversity of chemicals that could be stably dissolved in it, and the specifics of the planetary scenarios in which sulfuric acid oceans could accumulate. Ammonia is unlikely to occur as a liquid in its own right, and for that reason is less likely to be a solvent. Similarly, liquid hydrogen sulfide is unlikely to occur in bulk on a rocky planet. Liquid $CO_2$ and $SO_2$ remain candidates based on solubility criteria, but their likely cosmic abundances as liquids need to be explored further before they are considered. There is too little information on the solvation and chemical properties of liquid sulfur to draw clear conclusions. The cryogenic solvents methane, ethane and nitrogen are extremely implausible solvents for life due to their poor dissolution capability at the very low temperatures where they are liquid.

We briefly discussed the idea that life does not have to use a single solvent, but over geological time could adapt to use a different solvent as its environment changed. The concept of solvent replacement could be explored with in vitro evolution experiments.

Finally, we note that in our assessment of plausible solvents for life we have not included any criteria relating to the origin of life, we focused only on criteria related to life's ongoing chemistry. Origin of life studies have not defined either the environment nor the process by which terrestrial life might have originated – or rather, we have many scenarios but no way of reliably choosing between them. Absent any criteria for selecting a chemical path to life on Earth, where we know it happened, it is impractical to speculate how a non-terrestrial life in a non-aqueous solvent could originate. This would also be work for the future.

## *7 Acknowledgements.*

We are grateful to two anonymous reviewers for this detailed and constructive comments, especially one who pointed us to the potential of $H_2S$ as a solvent and Io as a home of sulfuric acid.

## *8 Author contributions*

Conceptualization: W.B., J.J.P., S.S.; methodology: W.B.; analysis W.B., J.J.P.; writing—original draft preparation, W.B.; writing—review and editing, W.B., J.J.P., S.S. All authors have read and agreed to the published version of the manuscript.

## *9 Conflict of interest*

The authors declare no conflict of interest.

## *10 Funding*

This research was partially supported by MIT.

## Supplementary Information

# Alternative solvents for life: framework for evaluation, current status and future research

William Bains[1,2,*], Janusz J. Petkowski[1,3,4], Sara Seager[1,5,6]

[1] Department of Earth, Atmospheric and Planetary Sciences, Massachusetts Institute of Technology, 77 Massachusetts Avenue, Cambridge, MA, 02139, USA

[2] School of Physics & Astronomy, Cardiff University, 4 The Parade, Cardiff CF24 3AA, UK.

[3] Faculty of Environmental Engineering, Wroclaw University of Science and Technology, 50-370 Wroclaw, Poland

[4] JJ Scientific, Mazowieckie, 02-792 Warsaw, Poland

[5] Department of Physics, Massachusetts Institute of Technology, 77 Massachusetts Avenue., Cambridge, MA 02139, USA

[6] Department of Aeronautics and Astronautics, Massachusetts Institute of Technology, 77 Massachusetts Avenue., Cambridge, MA 02139, USA

* correspondence: William Bains: bains@mit.edu

## 1. Volcanic gases

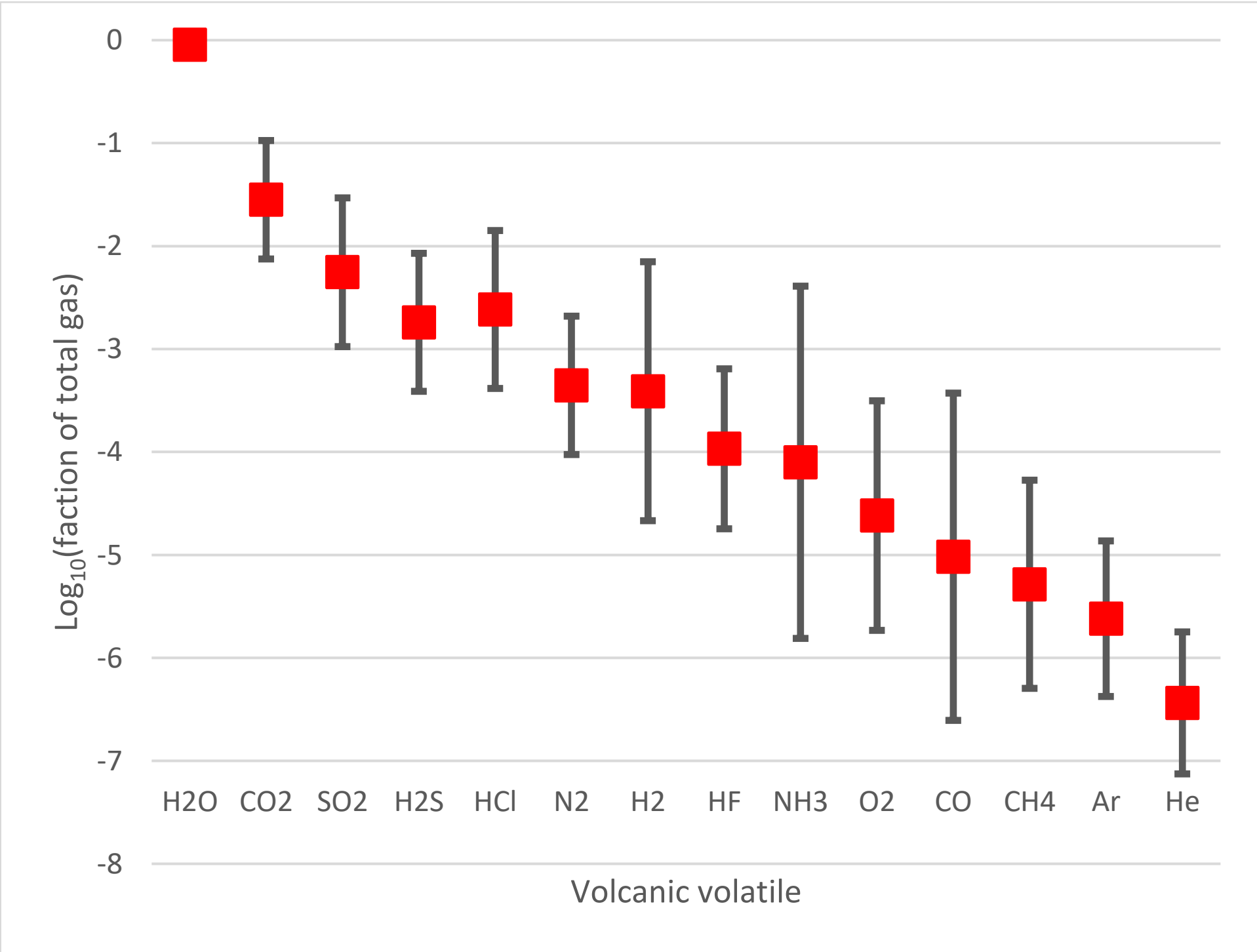

**Figure S1.** Relative abundance of volcanic gases on Earth. X axis: volcanic gases from volcanic and hydrothermal terrestrial systems on Earth, Y axis: abundance of each gas as a fraction of total released volatiles. Error bars show 1 standard deviation in measured fraction. Water is the dominant volcanic gas on Earth.

Water is by far the most abundant volcanic gas on Earth, followed by carbon dioxide. Figure S1 shows the relative abundance of volcanic gases from volcanic and hydrothermal terrestrial systems on Earth, as directly measured at the site. Gas abundance values are measured as a fraction of total volatiles in gases where a given gas species was measured. Only measurements taken at >120°C (i.e. where water was a vapour, and so solvent extraction into liquid water was not a concern) are considered. Data depicted on Figure S1 have been collected from (Giggenbach 1976, Gerlach 1980, Le Guern et al. 1982, Greenland 1984, Menyailov et al. 1986, Quisefit et al. 1989, Giggenbach et al. 1990, Giggenbach and Matsuo 1991, Poorter et al. 1991, Taran et al. 1991, Tedesco et al. 1991, Symonds et al. 1992, Taran et al. 1992, Chiodini et al. 1993, Taran et al. 1995, Fischer et al. 1998, Goff et al. 2000, Shinohara et al. 2002, Tassi et al. 2003a, Tassi et al. 2003b, Aiuppa et al. 2005, Clor et al. 2005, Lee et al. 2005, Liotta et al. 2010). Note that data collected on Figure S1 represents a wide range of geochemical context for gas emissions, and is meant to represent typical ranges of gas abundances, not any one particular volcanic chemistry. The standard deviation on the value for water is 0.0603, and is too small to show as an error bar on this graph.

The fraction of sulfur present in volcanic gases as hydrogen sulfide varies greatly depending on the gas temperature and the chemistry of the rocks from which $H_2S$ is formed and through which it passes on the way to the surface, as well as the amount of liquid water in the outgassing path (as $SO_2$ is much more soluble in water than $H_2S$). Figure S2 summarises the fraction of sulfur gases emitted as $H_2S$. The majority of volcanic vents produce more sulfur dioxide than hydrogen sulfide.

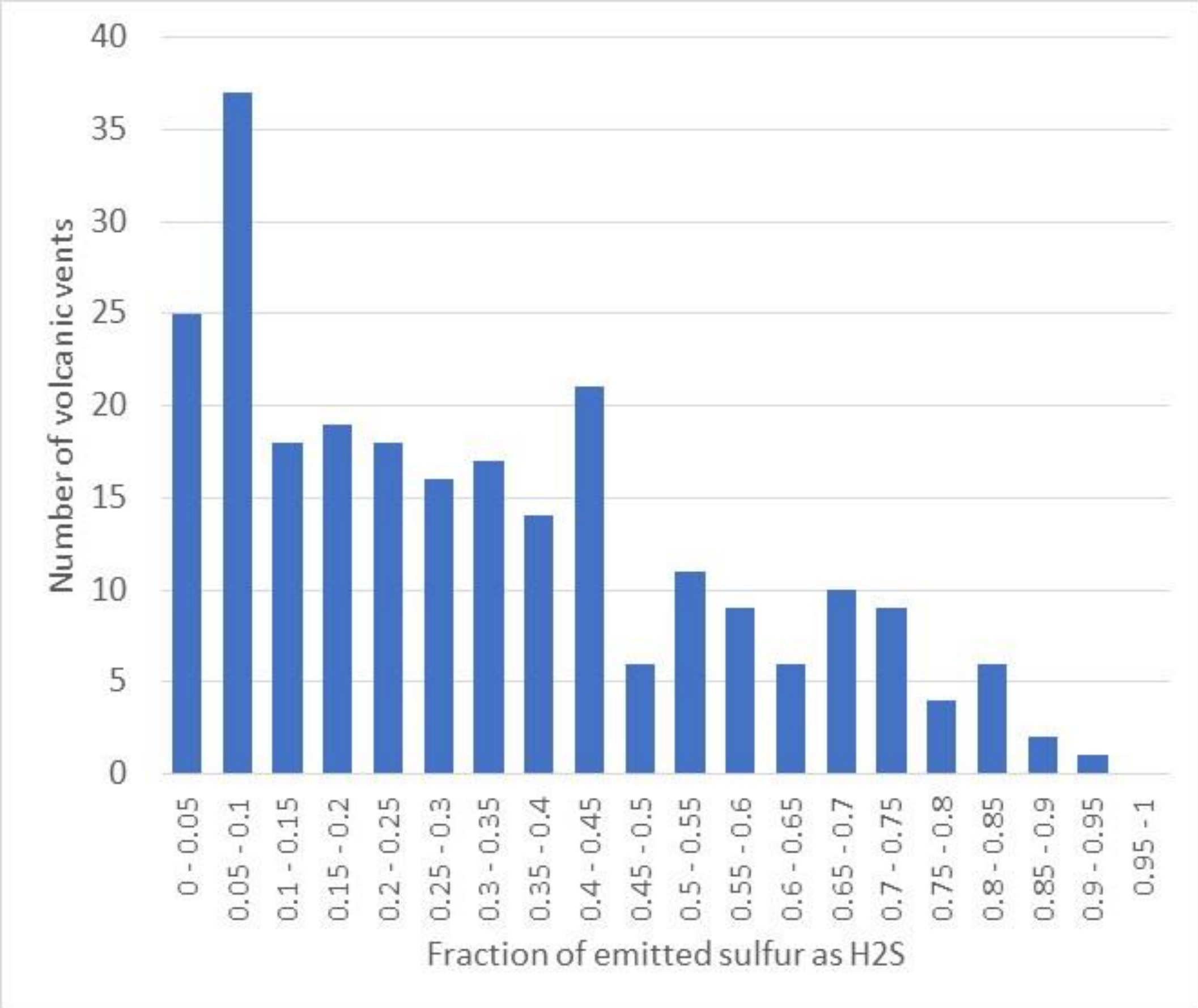

**Figure S2.** Fraction of sulfur-containing gases emitted as $H_2S$ from terrestrial volcanoes and fumaroles. References and vent selection are the same as for Figure S1. X axis: fraction of S gases that are $H_2S$, in bins of 0.05. Y axis: number of measurements with this ratio of gases. While $SO_2$ is the dominant gas in terrestrial volcanoes, some volcanic sources show high fractions of sulfur gases as $H_2S$.

## 2. Viscosity of solvents

What is the maximum and minimum viscosity of a solvent in which life can grow? The answer to this question requires us to estimate what the viscosity of the solvent inside a cell is. Such estimation is not straightforward. The cytoplasm behaves like a gel at large scales, and only has liquid-like behaviour at small scales (Golding and Cox 2006, Kwapiszewska et al. 2020). This is because the cytoplasm is not a dilute solution, but is packed with proteins and RNAs, and measuring viscosity by measuring movement of particles through the cytoplasm is in fact a combination of measurement of solvent viscosity and movement of a particle through a 3D lattice of macromolecules (Luby-Phelps 1994). Some estimates of the dynamic viscosity of the cytoplasm as measured on short length scales are provided in Table S1. These estimates show that on the smallest scales (i.e. nanometer) the cytoplasm does not have a substantially different viscosity from water for mesophilic organisms and mammalian cells. It is the smallest scales that concern us here; we assume that any non-terrestrial life will use polymers (Hoehler et al. 2020), and hence will have a highly structured, more viscous, cytoplasm on larger (nm) scales.

| Organism / system | Viscosity (Pa.s) | Method of estimation | Reference |
|---|---|---|---|
| Saccharomyces cerevisiae | 0.001 – 0.002 | NMR proton relaxation times for proteins | (Luby-Phelps 1999) |
| Bovine heart muscle | 0.0023 | NMR proton relaxation times for proteins | (Luby-Phelps 1999) |
| 3T3 fibroblasts | 0.001 – 0.002 | Rotational anisotropy of fluorescence from small molecule fluorescent probes | (Fushimi and Verkman 1991) |
| Mammalian cell lines | 0.0007 – 0.001 | Fluorescence energy transfer between mobile fluorophores | (Luby-Phelps et al. 1993) |
| Dormant fungal spores | 0.0035 – 0.015 | ESR of spin-labelled dissolved probe | (Dijksterhuis et al. 2007) |
| Germinating fungal spores | 0.0025 | ESR of spin-labelled dissolved probe | (Dijksterhuis et al. 2007) |
| Fibroblasts, cytoplasm adjacent to the cell membrane | 0.001 | Evanescent field fluorescence anisotropy | (Bicknese et al. 1993) |
| Pseudomonas aeruginosa | 0.0035 – 0.007 | Fluorescence anisotropy | (Cuecas et al. 2016) |
| E. coli | 0.007 | Diffusion of fluorescent proteins | (Elowitz et al. 1999) |
| Dictiostelium discoideum | 0.00189 – 0.00511 | Diffusion of GFP if actin networks are disrupted | (Potma et al. 2001) |

**Table S1.** Estimates of the dynamic viscosity of the cytoplasm in various types of cells.

Alternatively, we can ask what is the viscosity of the solvent inside cells, without dissolved proteins, organelles and other gel-forming materials. This is not pure water, as cells under some circumstances accumulate metabolites as protectants, which do not play any metabolic role. The cell in effect fills its interior with a substance other than water, and the bulk phase viscosity of that substance can be measured. We took this approach to estimate the viscosity of the solvent that life uses, and compared that to the viscosity of some of the solvents discussed in the main body of the paper.

The viscosity of water increases with decreasing temperature, and so the temperature range at which organisms can grow on Earth is an indication of the bulk viscosity that is consistent with terrestrial life. We know that terrestrial life can thrive in water at ~120°C, the highest temperature at which any terrestrial organism is known to grow (Takai et al. 2008). Water at 120°C has a dynamic viscosity of 0.0002291 Pa·s. This value provides an upper limit on the minimum viscosity. The only lower limit is set by theory, which suggests that extremely non-viscous fluids might hinder protein function (Branscomb and Russell 2019). The maximum viscosity for the solvent of life are provided by life that grows at low temperature and at low water activity.

Organisms have been reliably reported to grow (not merely survive) at temperatures down to -15°C (Junge et al. 2003, Mykytczuk et al. 2013, Panikov 2014, Mykytczuk et al. 2016). Contrary to common belief, the growth limit is rarely set by ice crystals forming inside cells exposed to slow cooling below 0°C, at least for unicellular organisms, as the internal fluid dehydrates and then vitrifies rather than freezing (Clarke et al. 2013). Freezing of the external solution can however cause toxic accumulation of solutes at the cell surface. Temperature limits for life include the availability of liquid water, the rate of metabolic reactions and the stability of proteins. Many proteins are 'marginally stable' at their normal functional temperature, with 3D-structural flexibility being a key part of their function (reviewed in (Goldenzweig and Fleishman 2018)). Lower temperature paradoxically renders these proteins too stable to function. As a consequence, many psychrophilic organisms accumulate chaotropic metabolites that destabilise protein structure as an adaptation to the cold (Chin et al. 2010). These include glycerol, fructose and trehalose, which is accumulated not as a carbon storage material but as a cryoprotectant (Wiemken 1990).

Xerophilic fungi can also accumulate up to 15% glycerol inside their cells in response to water stress at 25°C (Hocking and Norton 1983, Hocking 1986, Hallsworth and Magan 1995). (Note that metabolite concentrations are usually calculated as mg solute/gram wet weight, which includes the cell wall, internal vacuoles and potentially other structure not part of the cell's cytoplasm, so actual cytoplasmic concentration could be higher).

Thermophilic organisms can also modulate their sub-nm scale internal viscosity, increasing it with increasing temperature (presumably by synthesising compounds that 'thicken' the solvent) (Cuecas et al. 2016).

Table S2 lists the viscosities of the solvent phase (i.e. water + cryoprotectant or xeroprotectant only, not including other metabolites or proteins). Note that these are viscosities of the materials accumulated by organisms that have selectively synthesised or concentrated metabolites to allow growth at the selected temperature or water activity, so these values represent the optimum for growth in these conditions, not an extreme tolerated.

| **Organism** | **Temp. (°C)** | **Solvent milieu** | **Dynamic viscosity (Pa·s)** | **Comment** | **Ref. for organism** | **Ref. for viscosity** |
|---|---|---|---|---|---|---|
| Reference | 25°C | water | 0.0008811 | Included for comparison | | (Engineering Toolbox 2023) |
| Xerophilic fungus JW07JP13 | 10°C | 9.5% fructose | 0.01545 | | (Chin et al. 2010) | (Bulavin et al. 2017) † |
| Xerophilic fungus JW07JP13 | 30°C | 12.5% fructose | 0.01292 | | (Chin et al. 2010) | (Bulavin et al. 2017) † |
| Xerophilic fungus JW07JP13 | 1.7°C | 2% fructose | 0.00272 | | (Chin et al. 2010) | (Bulavin et al. 2017) † |
| Xerophilic fungus JW07JP13 | 10°C | 27.5% glycerol | 0.00329 | Under low water | (Chin et al. 2010) | (Segur and Oberstar 1951) |
| Xerophilic fungus JW07JP13 | 1.7°C | 3% glycerol | 0.001784 | Under low water | (Chin et al. 2010) | (Segur and Oberstar 1951) |
| Unspecified bacteria | -10ºC | Water | 0.002685 | Supercooled cloud droplets | (Sattler et al. 2001) | (Dehaoui et al. 2015) |
| Eurotium ehevalieri Mangin FRR 547 | 25ºC | 12% glycerol | 0.001185 | | (Hocking 1986) | (Segur and Oberstar 1951) |
| Saccharomyces cereviseae | 4ºC | 4% glycerol | 0.001760 | On shifting from 20ºC | (Panadero et al. 2006) | (Segur and Oberstar 1951) |
| Saccharomyces cereviseae | 0ºC | 3.4% trehalose | 0.001993 | Author's units are inconsistent | (Kandror et al. 2004) | (Sampedro et al. 2002) |
| Hebeloma sp | 0ºC | 2.5% trehalose | 0.001968 | | (Tibbett et al. 2002) | (Sampedro et al. 2002) |
| Phlebiopsis gigantea | 24ºC | 4.5% trehalose | 0.001295 | | (Ianutsevich et al. 2023) | (Sampedro et al. 2002) |

**Table S2.** An overview of the viscosities of the solvent phase. † The viscosity of glucose and fructose are very similar (Flood and Puagsa 2000), and as the absolute viscosity of fructose was not available, only 'relative viscosity', the viscosity of glucose was used as a surrogate in these measures.

The different viscosities are plotted in Figure S3. Also plotted is the viscosity of sulfuric acid, liquid sulfur and liquid $CO_2$, as a function of temperature.

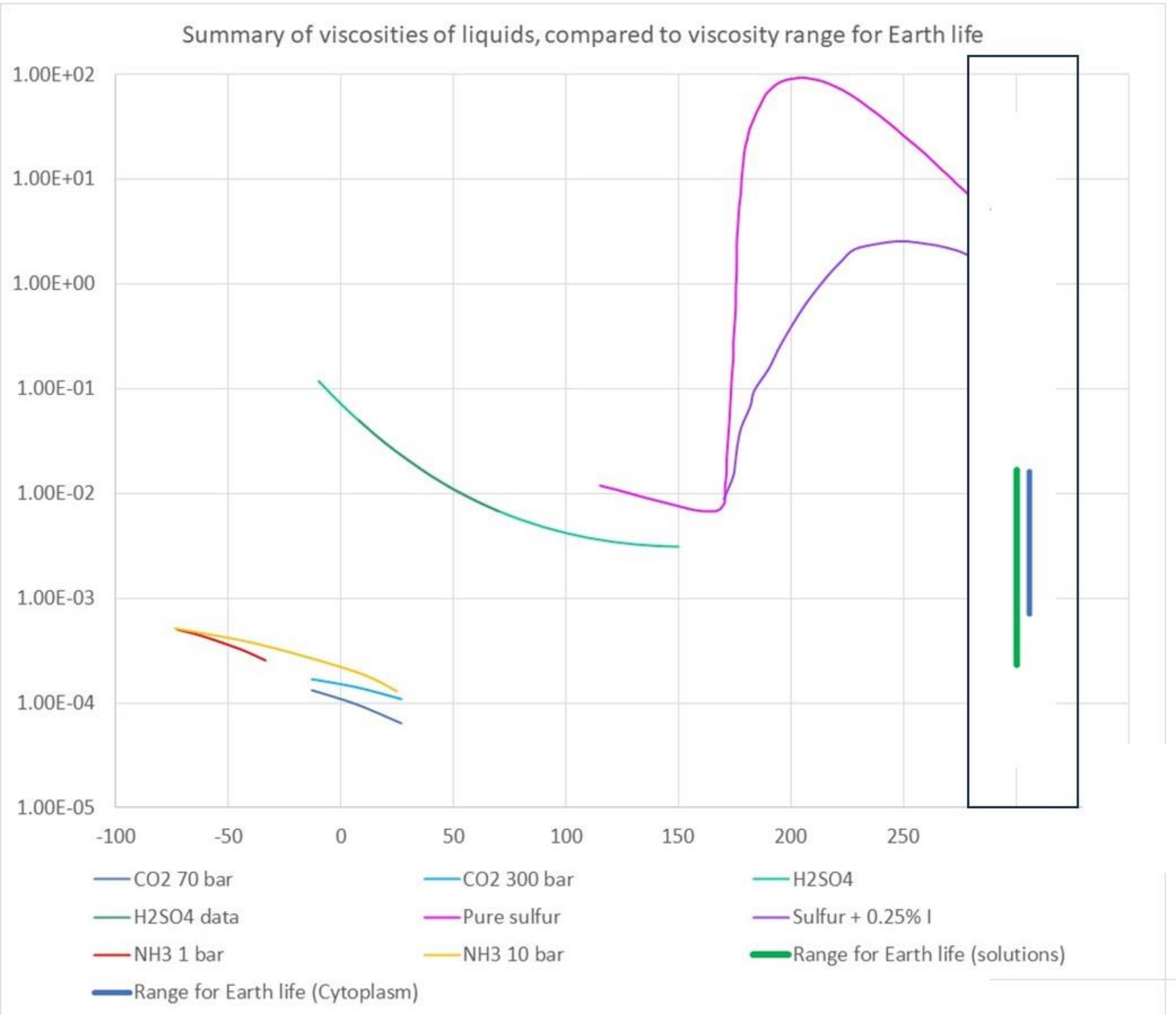

**Figure S3.** Viscosity of solvents compared to the viscosity of the interior of cells. X axis: temperature (°C), Y axis: viscosity (Pa·s) on log scale. Sulfuric acid and liquid sulfur have viscosities falling inside those found for water solutions in Earth life at some temperatures. Liquid carbon dioxide has a lower viscosity than any viscosity value experienced by Earth life.

Figure S3 shows viscosities of liquid $CO_2$ (at 70 bar and 300 bar), liquid sulfuric acid (measured for 100% w/w and extrapolated for 98% w/w), liquid sulfur (for pure sulfur and for sulfur containing 0.25% elemental iodine) and liquid ammonia at 1 bar and 10 bar. Also shown are ranges for the measured viscosity of the cytoplasm (green bar) from Table S1 above, and the predicted viscosity of solute-loaded cytoplasm (blue bar) from Table S2 above. Viscosity of sulfuric acid 100% w/w sulfuric acid is provided by (Greenwood and Thompson 1959). The viscosity of 95% w/w at 25°C is only 11% lower than 100% w/w sulfuric acid (Liler 2012), suggesting that the viscosity value for 95% w/w sulfuric acid at 25 °C is a good proxy for all highly concentrated sulfuric acids. Figure S3 shows measured viscosity values and values extrapolated from measurements using the equation

$$\eta = e^{3.1939 \cdot 10^{-7} . T^3 + 2.2835 \cdot 10^{-4} \cdot T^2 - 4.7913 \cdot 10^{-2} \cdot T - 2.6487} \quad (1)$$

where T is the absolute temperature and η is the dynamic viscosity in Pa·s. Note that the extrapolated values below 7.5°C do not apply to pure sulfuric acid, as $H_2SO_4$ solidifies at 7.5°C. The viscosity of liquid sulfur is from (Powell and Eyring 1943). The viscosity of liquid $CO_2$ is from (Padua et al. 1994). The viscosity of liquid ammonia is from (Engineering Toolbox 2023)

Our conclusion is that viscosity is not a barrier to using sulfuric acid above ~30°C, liquid sulfur below ~170°C, or liquid ammonia below -0°C depending on pressure. Liquid carbon dioxide lies below the viscosity of the most fluid water known to support life. However, emphasize that it is not known if lower or higher viscosity than water is a insurmountable barrier for life or not.

## 3. Examples of differential effects of sulfuric acid concentration on solvolysis

The rate of reactions in sulfuric acid depend on the concentration of the acid for three reasons. First, high acidity means that acid-catalysed reactions usually go faster with increasing concentration. This includes reactions that rely on protonation of very weakly basic species such as carbonyl or phenyl groups. Second, the presence of $SO_3$ or equivalent species mean that sulfonation reactions can happen in concentrated acid that do not happen in dilute acid. However, reactions that are limited by the concentration of water of $OH^-$ ions go more slowly in concentrated acid, as water is both present in lower amounts and is chemically bound to $H_2SO_4$ molecules, and $OH^-$ is effectively absent in concentrated acid.

Figure S4 and Figure S5 give examples of reaction rates as a function of acid concentration and show that reaction rate can increase with increasing acid (Compound D), decrease with increasing acid (compound A), or show more complicated patterns of increase and decrease with acid concentration (compounds B and C). Even the reaction rate of apparently similar compounds like C and D can have dramatically different kinetic responses to changes in acid concentration. See (Bains et al. 2021a, Bains et al. 2021b) for a more thorough review of sulfuric acid solvolysis kinetics.

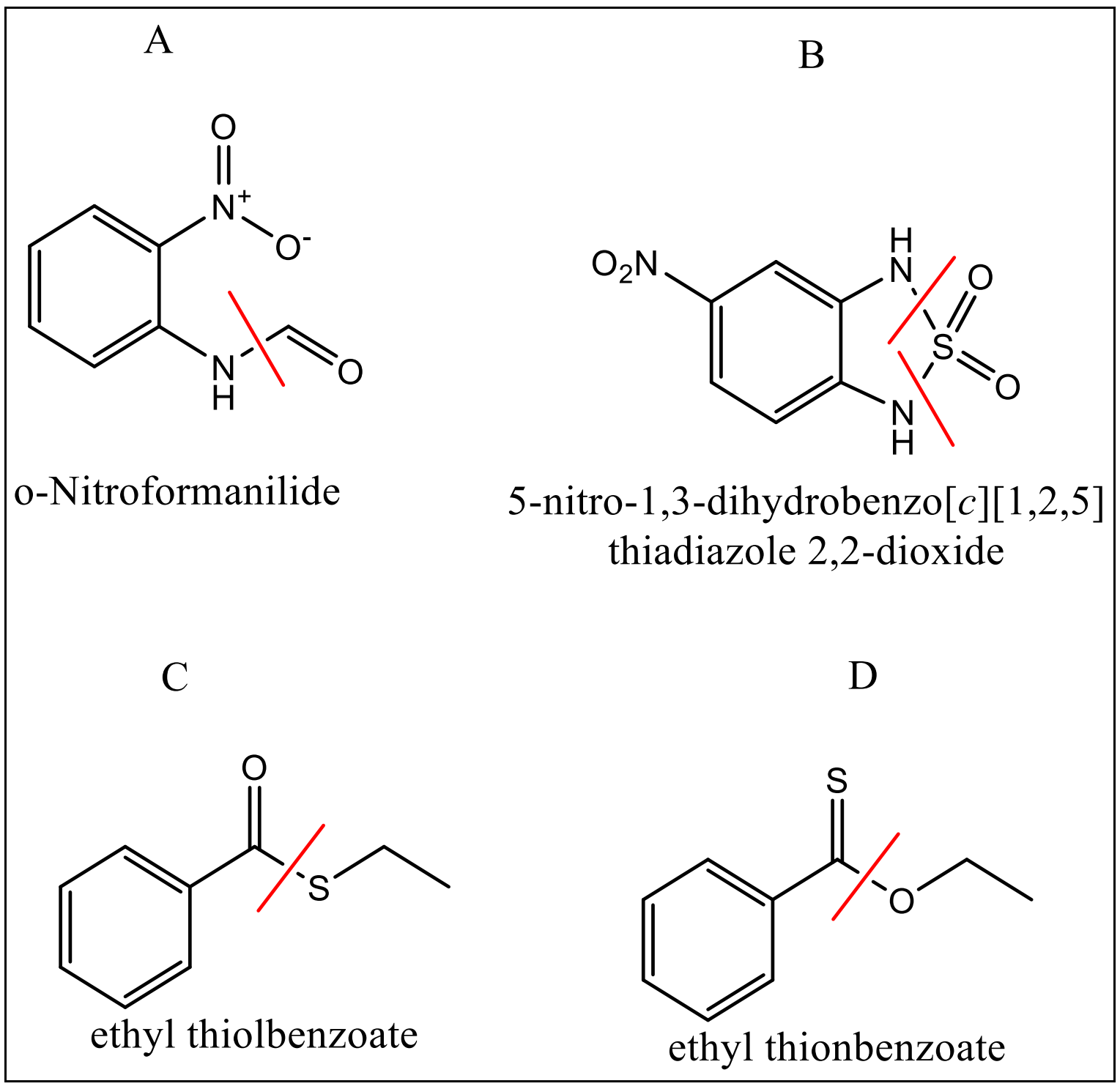

**Figure S4.** Structures of compounds A-D. The rate of reaction of compounds A-D in sulfuric acid is plotted in Figure S5. Red lines show where the compounds are hydrolysed in sulfuric acid.

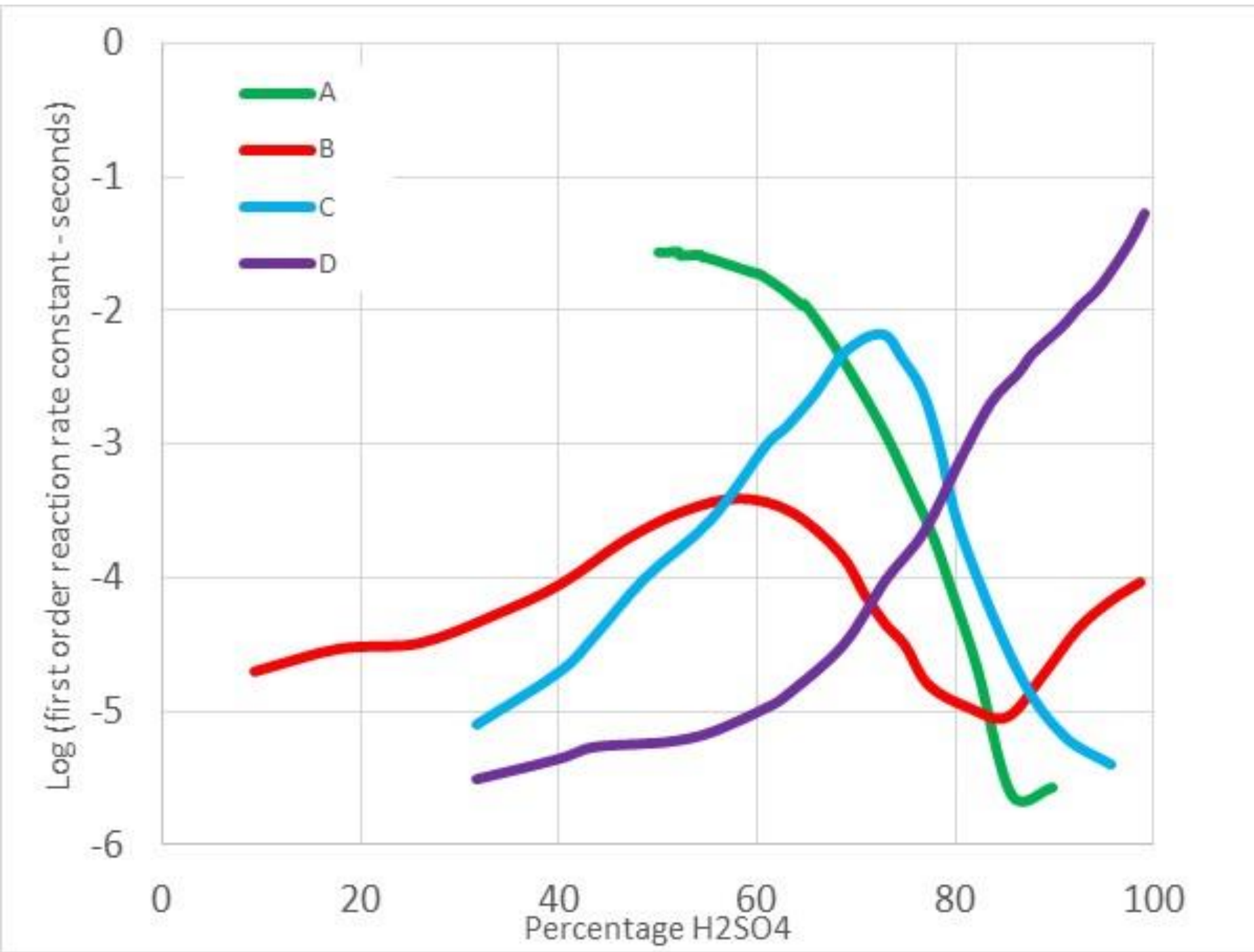

**Figure S5.** Different chemical structures (compounds A-D depicted in Figure S4) show very different reaction rates as a function of sulfuric acid concentration. X axis: sulfuric acid concentration (% w/w). Y axis – $\log_{10}$ of first order reaction rate ($sec^{-1}$). Plotted is the experimental data, not a model. Data from: A - (Vinnik et al. 1976), B - (Gediz Erturk and Bekdemir 2015), C and D - (Edward and Wong 1977).

## 4. Carbon dioxide density and solubility

Solubility of molecules in supercritical $CO_2$ is empirically found to be related to the fluid's density by the following equation:

$$\ln(s) = A + B \cdot T^2 + C \cdot T + D \cdot \rho \quad (2)$$

where A - D are constants, T is the absolute temperature and ρ is the density (Del Valle and Aguilera 1988). The constant D is related to the average number of molecules in the solvated complex, which is at first approximation a measure of the size of the solute molecule. The reason for this is that to solvate a molecule the solvent needs to be packed densely round that molecule as a solvation shell. If the density of the liquid is lower than the density of a solvation shell then solvation is disfavoured. For example, for esters in vegetable oil, A was found to be -7.59498, B=7.0309.10$^{-5}$, C = 0.040703, and D=0.012089 from the data in (Del Valle and Aguilera 1988).

The density of supercritical carbon dioxide changes quite rapidly with changes in temperature at near-critical pressures, and with pressure and near critical temperatures. Figure S6 shows the phase diagram for $CO_2$ in terms of density

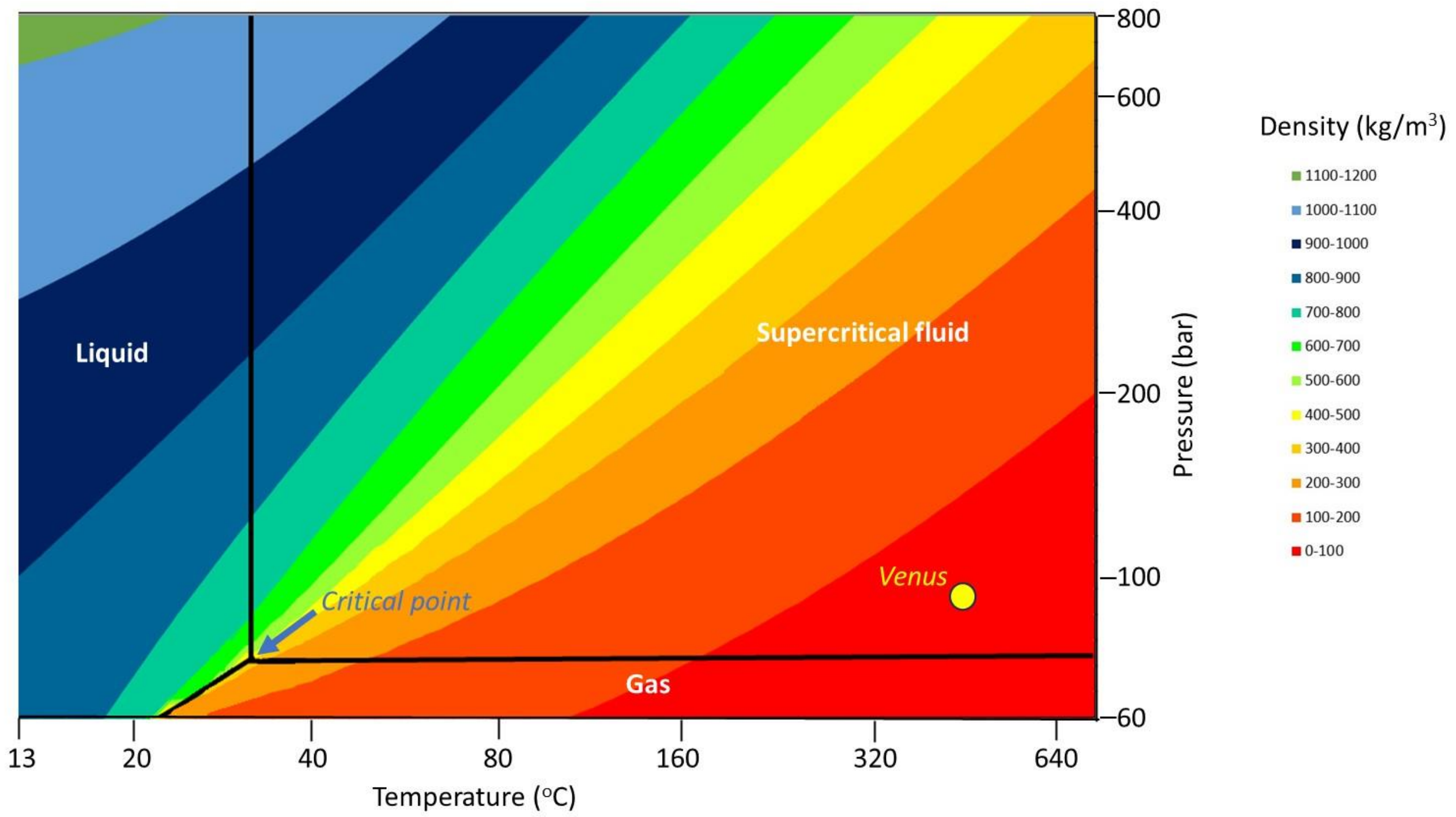

**Figure S6.** Density of $CO_2$ as a function of pressure and temperature. X axis: temperature (°C). Y axis: Pressure (bar). Colour scale shows the density of the fluid. Also plotted are the phase boundaries. Above the critical temperature and pressure (the critical point) the material is a supercritical fluid, that cannot be compressed to a liquid. Blue colours are regions of the T/P space where the density is over the ~800 kg/m$^3$ needed to dissolve proteins. The T/P of Venus' surface atmosphere is indicated by the yellow circle.

From equation 2 and the density/pressure/temperature diagram on Figure S6, we can ask what effect a change in temperature has on the solubility at constant pressure for palm oil, as an example (Figure S7).

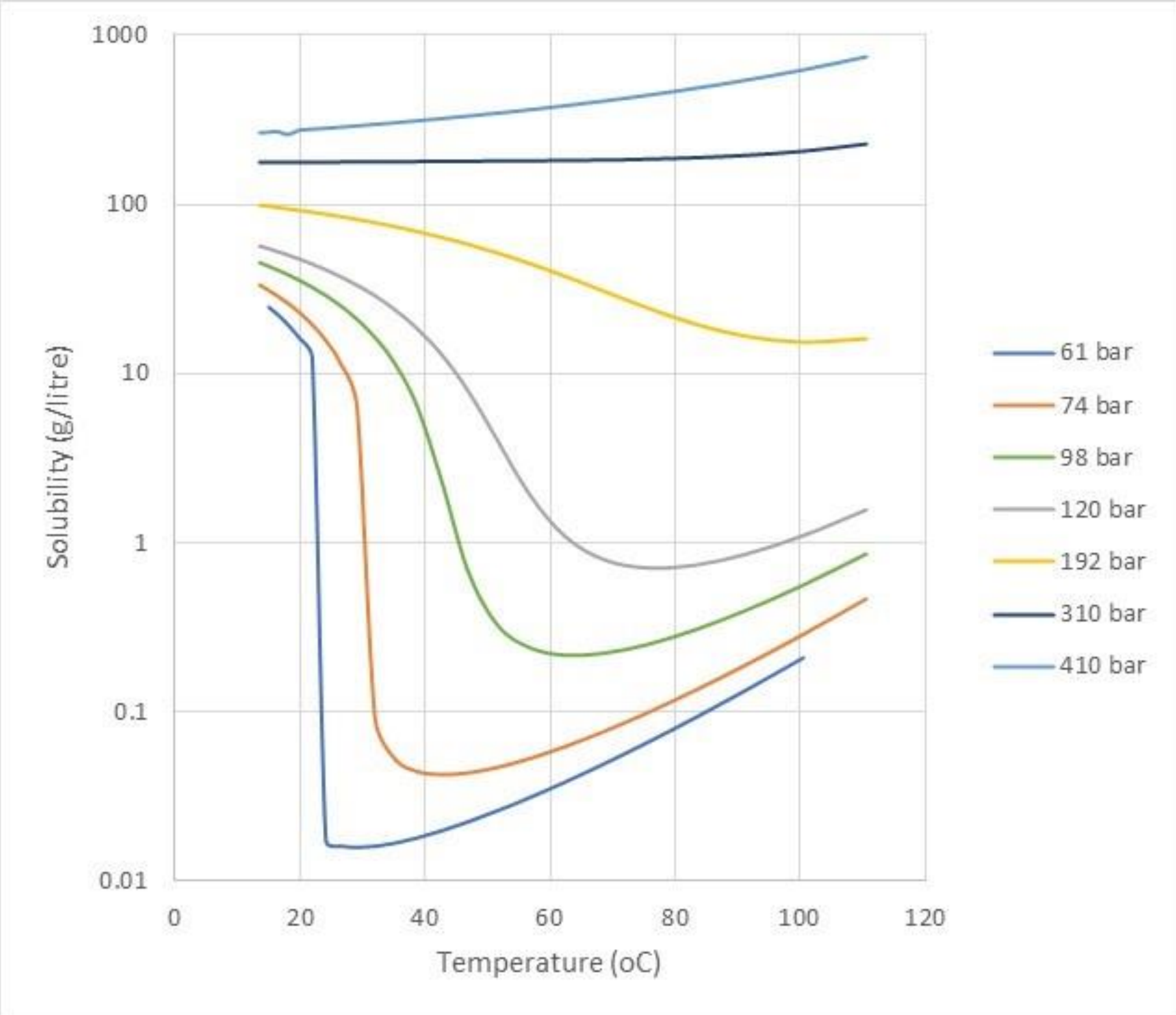

**Figure S7.** Solubility of palm oils in supercritical $CO_2$ as a function of temperature for constant pressure. X axis: Temperature (°C), Y axis: solubility (g/l). See SI text for details.

The curve for 61 bar on Figure S7 shows a dramatic drop-off of solubility at 24 degrees, as 61 bar is below the critical pressure for $CO_2$, and so this represents the change in solubility as liquid $CO_2$ (below 24°C) boils into gas (above 24°C). The other curves show the change in solubility as the fluid moves from the liquid phase below 31.5°C to the supercritical phase. Note that even for 90 bar (slightly higher than the surface pressure on Venus) there is a dramatic change in solubility between 40°C and 60°C, a 20°C change: Earth land surface temperatures often differ by more degrees between day and night. Only at much higher pressures the solubility is relatively constant with temperature.

## 5. Ionic liquids

Ionic liquids are ionic compounds that have a very low melting point, usually defined as below 100°C. (Venkatraman et al. 2018) list 1330 ionic liquids with a melting point below 100°C, 249 with a meting point below 0°C. The structures of a few of such liquids are depicted on Figure S8.

[1]

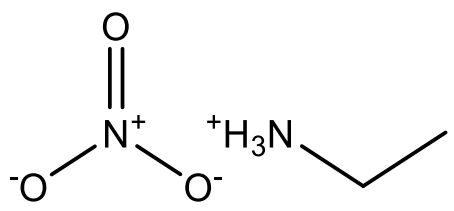

ethylammonium nitrate

[2]

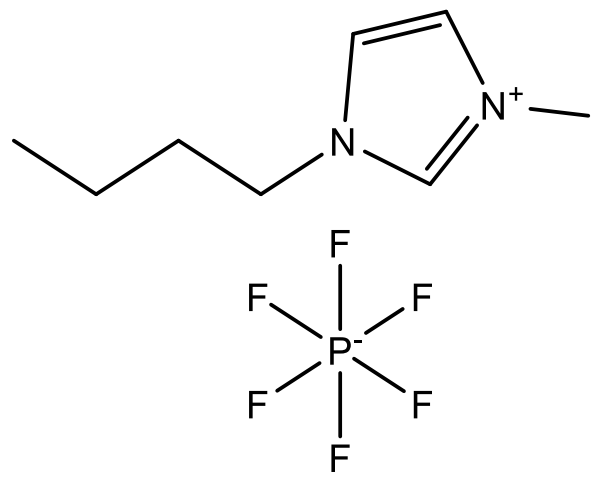

1-butyl-3-methylimidazolium hexafluorophosphate

[3]

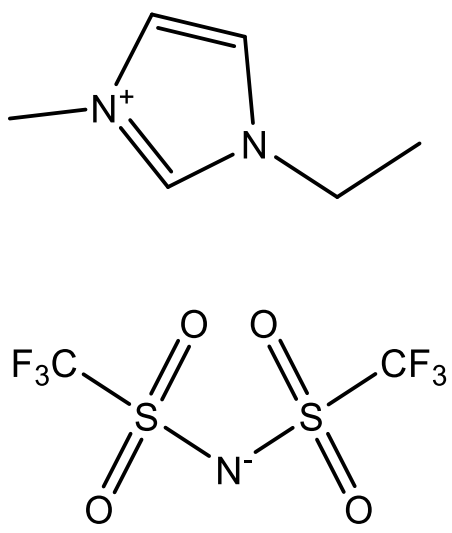

1-ethyl-3-methylimidazolium bis-(trifluoromethanesulfonyl) imide

[4]

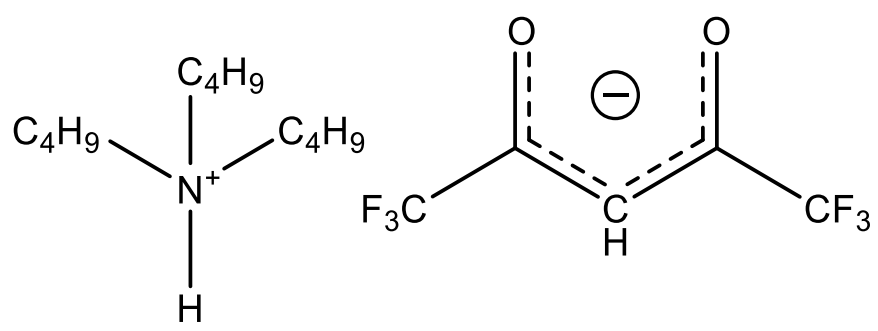

tributylammonium 1,3-di-(trifluormethyl)-pentane-2,4-diketonate

**Figure S8.** Examples of ionic liquids. Compound 1 (Walden 1914) is the first ionic liquid ever reported, compound 2 is one of the most widely used ionic liquids (Hajipour and Rafiee 2009). Compound 3 is an example of an ionic liquid that is stable to water and air (Endres and Zein El Abedin 2006). Compound 4 has one of the lowest known melting points for an ionic liquid, melting at -96°C (Gupta et al. 2004).

## 6. Relative stability of methane and methanol

As an illustration of the relative reactivity of methane and methanol, we calculated the reaction rate constants for the reaction of methanol and methane with ·O and ·OH radicals (the primary reactive species in an $N_2/CO_2/H_2O$ atmosphere under photochemistry), based on the kinetic data in (NIST 2024). The geometric mean of the experimentally determined values of the constants A and Ea in the Arrhenius equation were calculated for each reaction of $CH_4$ or $CH_3OH$ with ·O or ·OH. The rate constant for that reaction at a given temperature was then calculated using the Arrhenius equation

$$k = A.e^{-\frac{Ea}{R.T}}$$

where R is the gas constant 8.314 J/mol·K and T is the absolute temperature. The rate constants are plotted in Figure S9. The reaction rate will depend on the relative abundance of ·O and ·OH radicals, which will itself depend on the composition and illumination of the atmosphere, but in all cases the rate of destruction of methanol will be several orders of magnitude faster than that of methane, especially at temperatures below 200K postulated by (Tang et al. 2006)

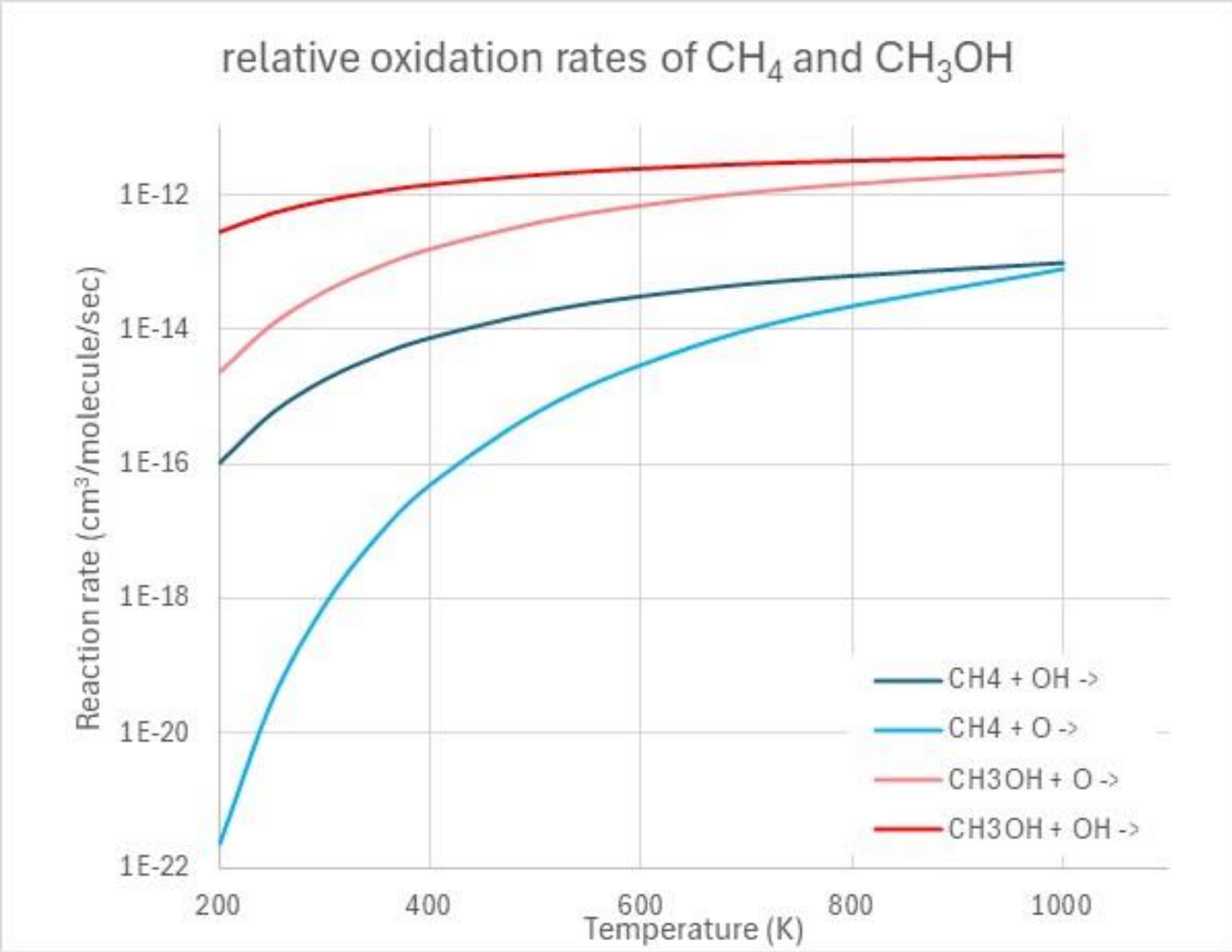

**Figure S9.** Rate constants for the reaction of methane and methanol with ·O and ·OH. At all temperatures the reaction of methane with either species is slower than the reaction of methanol with that species. Reactions are from the reacting species to any of several products.